\documentclass[fleqn,usenatbib]{mnras}

\usepackage[T1]{fontenc}
\usepackage{ae,aecompl}

\usepackage{graphicx}	
\usepackage{amsmath}	
\usepackage{amssymb}	
\usepackage{breqn}
\usepackage{color}
\usepackage{cases}
\usepackage{enumerate}
\usepackage{threeparttable}
\usepackage[multidot]{grffile}

\usepackage{txfonts}
\hypersetup{draft}

\usepackage[authoryear]{natbib}
\bibpunct{(}{)}{;}{a}{}{,}
\newcommand\meraxes{{\sc Meraxes}~}
\newcommand\meraxesns{{\sc Meraxes}}
\defcitealias{Qin2017}{Q17}
\defcitealias{Liu2016}{L16}

\title[Sizes, angular momenta and morphologies at high-z]{Dark-ages Reionization and Galaxy Formation Simulation -- XVII. \\ Sizes, angular momenta and morphologies of high redshift galaxies}

\author[M. A. Marshall et al.]{Madeline A. Marshall$^{1,2}$\thanks{E-mail: madelinem1@student.unimelb.edu.au (MAM); swyithe@unimelb.edu.au (JSBW)}, Simon J. Mutch$^{1,2}$, Yuxiang Qin$^{1,3}$, Gregory B. Poole$^4$,
\newauthor J. Stuart B. Wyithe$^{1,2}$\footnotemark[1]
\\
$^{1}$ School of Physics, University of Melbourne, Parkville, VIC 3010, Australia\\
$^{2}$ ARC Centre of Excellence for All Sky Astrophysics in 3 Dimensions (ASTRO 3D)\\
$^{3}$ Scuola Normale Superiore, Piazza dei Cavalieri 7, I-56126 Pisa, PI, Italy\\
$^{4}$ Centre for Astrophysics and Supercomputing, Swinburne University of Technology, PO Box 218, Hawthorn VIC 3122, Australia\\}

\date{Accepted XXX. Received YYY; in original form ZZZ}

\pubyear{2019}

\begin{document}
\label{firstpage}
\pagerange{\pageref{firstpage}--\pageref{lastpage}}
\maketitle

\begin{abstract}
We study the sizes, angular momenta and morphologies of high-redshift galaxies using an update of the \meraxes semi-analytic galaxy evolution model.
Our model successfully reproduces a range of observations from redshifts $z=0$--10.
We find that the effective radius of a galaxy disc scales with UV luminosity as $R_e\propto L_{\textrm{UV}}^{0.33}$ at $z=5$--10, and with stellar mass as $R_e\propto M_\ast^{0.24}$ at $z=5$ but with a slope that increases at higher redshifts.
Our model predicts that the median galaxy size scales with redshift as $R_e \propto (1+z)^{-m}$, where  $m=1.98\pm0.07$ for galaxies with (0.3--1)$L^\ast_{z=3}$ and $m=2.15\pm0.05$ for galaxies with (0.12--0.3)$L^\ast_{z=3}$.
We find that the ratio between stellar and halo specific angular momentum is typically less than one and decreases with halo and stellar mass.
This relation shows no redshift dependence, while the relation between specific angular momentum and stellar mass decreases by $\sim0.5$ dex from $z=7$ to $z=2$.
Our model reproduces the distribution of local galaxy morphologies, with bulges formed predominantly through galaxy mergers for low-mass galaxies, disc-instabilities for galaxies with $M_\ast\simeq10^{10}$--$10^{11.5}M_\odot$, and major mergers for the most massive galaxies. At high redshifts, we find galaxy morphologies that are predominantly bulge-dominated.
\end{abstract}
\begin{keywords}
galaxies: evolution -- galaxies: bulges -- galaxies: high-redshift.
\end{keywords}



\section{Introduction}
Size, angular momentum and morphology are three of the most fundamental galaxy properties, and are integral probes for understanding the structure and growth of galaxies. Observing these properties at high redshifts offers an invaluable insight into galaxy formation and evolution in the early Universe.

The sizes of high-redshift Lyman-break galaxies have been measured using deep Hubble Space Telescope (HST) fields in a number of studies from $z\sim6$--12 \citep[e.g.][]{Oesch2010,Mosleh2012,Grazian2012,Ono2013,Huang2013,Holwerda2015,Kawamata2015,Shibuya2015,Kawamata2018}. 
They find sizes consistent with an evolution of the galaxy effective radius $R_e \propto (1+z)^{-m}$ at fixed luminosity, with measurements of $m$ typically in the range of $1\lesssim m \lesssim 1.5$.

While observing galaxy morphologies at high-redshift is challenging \citep[see e.g.][]{Abraham2001}, observations generally find that high-redshift galaxies are more clumpy and irregular than those at low redshift \citep{Abraham2001,Papovich2005,Elmegreen2014}. Several studies find that there is a larger fraction of spheroids at higher redshifts \citep[e.g.][]{Bundy2005,Franceschini2006,Ravindranath2006,Lotz2006,Dahlen2007}, though \citet{Ravindranath2006} find that the $z\sim2.5$--5 population is not dominated by spheroids but by extended disc-like galaxies and irregulars or merger-like systems.

Unfortunately, there are only a few observational studies of galaxy angular momentum at redshifts $z>1$ \citep[e.g.][]{Swinbank2017,Alcorn2018,Okamura2018}, since obtaining large spectroscopic samples to measure kinematics and thus specific angular momenta at high redshifts is difficult. \citet{Okamura2018}, for example, estimated the specific angular momentum of $z\sim2$, 3 and 4 galaxies from their measured disc sizes, using the analytic model of \citet{Mo1998}, and found little redshift evolution of the ratio of stellar to halo specific angular momentum.
Because of the difficulty of observing these properties at high redshift, models and simulations play a necessary role for measuring and understanding their evolution in the early Universe.

In the hierarchical structure formation scenario, gas cools in the centres of dark matter haloes to form galaxies. 
The \citet{Mo1998} analytical model of this process predicts that if galaxies are thin exponential discs with flat rotation curves, and specific angular momentum is conserved, the scale length of a galaxy disc $R_s$ is 
\begin{equation}
R_{s}=\frac{\lambda}{\sqrt{2}}\left(\frac{J_d}{J_H}\frac{M_{\textrm{vir}}}{M_d}\right)R_{\textrm{vir}}
\label{eq:MoSize}
\end{equation}
where $J_d$ is its angular momentum, $M_d$ its mass, $R_{\textrm{vir}}$, $M_{\textrm{vir}}$, $J_H$ and $j_H$ are the virial radius, virial mass, and total and specific angular momentum of its dark matter halo, and $\lambda=|\textbf{j}_H|/(\sqrt{2}R_{\textrm{vir}}V_{\textrm{vir}})$ is the halo spin parameter (\citealt{Bullock2001}; see \citealt{Angel2016} for an examination at high redshift). Note that this model is an extension of the \citet{Fall1983} model which assumed $R_{s}=\lambda R_{\textrm{vir}} / \sqrt{2}$, with $J_d=J_H$ and $M_d=M_{\textrm{vir}}$.
The virial radius of a dark matter halo is
\begin{equation}
R_{\textrm{vir}}=\left(\frac{GM_{\textrm{vir}}}{100H^2(z)}\right)^{1/3}=\frac{V_{\textrm{vir}}}{10H(z)},
\end{equation}
where  $V_{\textrm{vir}}$ is the virial velocity, and $H(z)$ is the Hubble parameter, which is proportional to $(1+z)^{1.5}$ at high redshifts \citep{Carroll1992}. Hence, the simple model of \citet{Mo1998} predicts that sizes of discs scale with redshift as $(1+z)^{-1.5}$ at fixed circular velocity, or $(1+z)^{-1}$ at fixed halo mass.
Measurements of $m$ typically lie between these two values \citep[e.g.][]{Bouwens2004,Oesch2010,Ono2013,Kawamata2015,Shibuya2015,Laporte2016,Kawamata2018}.

Semi-analytic models often use the \citet{Fall1983} model and assume $R_{s}=\lambda R_{\textrm{vir}} / \sqrt{2}$ to predict the sizes of galaxies within their simulated dark matter haloes \citep[e.g.][]{Croton2006a,Mutch2016,Liu2016}. While simplistic, this model has had success in reproducing the size evolution of galaxies from $z\sim5$--$9$ \citep{Liu2016}.
More advanced semi-analytic models improve their disc size estimates by explicitly tracking the angular momentum of galaxy discs \citep[e.g.][]{Lagos2009,Guo2011,Benson2012,Stevens2016,Tonini2016,Xie2017,Zoldan2019}. 
Under the assumption that galaxy discs have a constant velocity profile and an exponential surface density profile, their specific angular momentum is $j=2R_sV$, and so their radii can be estimated as $R_s=j/2V$. These models generally assume that the velocity of the galaxy disc $V$ is equal to the maximum circular velocity of the surrounding dark matter halo \citep[e.g.][]{Guo2011,Tonini2016,Zoldan2019}.
\citet{Stevens2016} introduce a more sophisticated semi-analytic model, splitting galaxy discs into a series of constant $j$ annuli, which allows the model to self-consistently evolve the radial and angular momentum structure of galaxy discs.
The \citet{Stevens2016} model, alongside hydrodynamical simulations \citep[e.g.][]{Pedrosa2015,Teklu2015,Lagos2017}, have made predictions for the redshift evolution of angular momentum, with some finding no or minimal increase of $j$ with decreasing redshifts, while others support the gradual growth of $j$ in galaxies. However, at high redshifts ($z>2$), no theoretical studies have been compared with the (minimal) observational angular momentum data.

To track morphology, semi-analytic models often assume that bulges can be formed through mergers and disc instabilities, and are able to get good agreement with low-redshift morphological observations \citep[e.g.][]{Guo2011,Tonini2016,
IzquierdoVillalba2019}.
Cosmological hydrodynamical simulations have also recently been able to make reasonable predictions for galaxy morphologies at low-redshifts \citep[e.g.][]{Snyder2015,Correa2017,Clauwens2018,RodriguezGomez2019}.
Investigating higher redshifts, \citet{Trayford2018} and \citet{Pillepich2019} both find that the fraction of galaxies with disc-like morphologies increases from $z\simeq6$--0.5, with \citet{Trayford2018} predicting that galaxies at $z>1.5$ become disturbed, with the Hubble sequence dominating at $z<1.5$.

\meraxes \citep{Mutch2016} is a semi-analytic galaxy formation model designed to study high-redshift galaxy evolution. In \citet{Liu2016} (hereafter \citetalias{Liu2016}), \meraxes was used to investigate the sizes of high-redshift galaxies. This model assumed $R_s=R_{\textrm{vir}}\lambda/\sqrt{2}$, included no angular momentum tracking, and assumed that all galaxies were exponential discs for simplicity.
In this paper, we use \meraxes to study the evolution of galaxy sizes, angular momentum and morphology.
We implement a new method to determine galaxy sizes and velocities, introduce angular momentum tracking, and implement a bulge growth model to allow for morphological studies. 

This paper is organised as follows. In Section \ref{Meraxes} we introduce \meraxes and detail these new model updates, including a model for bulge formation and growth, a density-dependent star formation prescription, and radius and angular momentum tracking. We calibrate the model and verify its predictions for low redshift sizes, angular momenta and morphologies in Section \ref{Verification}. We then make predictions for the high-redshift evolution of galaxy sizes in Section \ref{HighZSizes}, angular momenta in Section \ref{HighZAngularMomentum} and morphologies in Section  \ref{HighZMorphology} . We conclude in Section  \ref{Conclusion}.

Throughout this work, we adopt the \citet{Planck2015} cosmological parameters: ($h, ~\Omega_m, ~\Omega_b, ~\Omega_\Lambda, \sigma_8, n_s$)=(0.678, 0.308, 0.0484, 0.692, 0.815, 0.968). All magnitudes are presented in the AB system \citep{Oke1983}.

\section{Semi-analytic model}
\label{Meraxes}
In this section we describe additions to the semi-analytic model \meraxes that allow calculation of the disc scale length, bulge mass, and angular momentum of galaxies in the model. Readers interested in the calibration and results may skip to Section \ref{Verification}.

\subsection{N-body simulations}
In this work, we apply \meraxes to two collisionless N-body dark matter simulations, \textit{Tiamat} and \textit{Tiamat-125-HR}.
The \textit{Tiamat} simulation \citep{Poole2016,Poole2017} has a co-moving box size of $(67.8h^{-1}\textrm{Mpc})^3$, and contains $2160^3$ particles each of mass $2.64\times10^6 h^{-1}M_\odot$. \textit{Tiamat} outputs 164 snapshots from $z=35$ to $z=1.8$, with a temporal resolution of 11.1 Myr to $z=5$, and with the snapshots between $z = 5$ and $z=1.8$ separated equally in units of Hubble time.
The high mass and temporal resolution of \textit{Tiamat} make it an extremely accurate and ideal simulation for studying galaxy evolution at high-redshifts. \textit{Tiamat} is used throughout this work unless otherwise specified.

To investigate lower redshifts, we run \meraxes on \textit{Tiamat-125-HR} \citep{Poole2017}, a low-redshift compliment to \textit{Tiamat}, with the same cosmology and snapshot cadence extended to $z=0$. \textit{Tiamat-125-HR} has a larger box size of $(125h^{-1}\textrm{Mpc})^3$ and a lower mass resolution, with $1080^3$ particles with mass $1.33\times10^8 h^{-1}M_\odot$, adequate for investigating low-redshift galaxy evolution.
For a detailed description of these simulations, see \citet{Poole2016} and \citet{Poole2017}.

\subsection{\meraxes}
\meraxes \citep{Mutch2016} is a semi-analytic model specifically designed to study galaxy evolution during the Epoch of Reionization. 
Using the properties of the dark matter haloes from \textit{Tiamat} and \textit{Tiamat-125-HR}, \meraxes models the baryonic physics involved in galaxy formation and evolution analytically, including prescriptions for physical processes such as gas cooling, star formation, AGN and supernova feedback, and black hole growth. The various free parameters of the model are calibrated to reproduce certain observations, such as the evolution of the galaxy stellar mass function, and, for the parameters related to reionisation, measurements of the integrated free electron Thomson scattering optical depth; we discuss our updated calibration step in detail in Section \ref{sec:Calibration}. The model outputs a large set of properties describing each simulated galaxy, such as the mass of hot and cold gas, stars, and the central black hole, its star formation rate, disc size, and type (central or satellite), at each snapshot of the dark matter simulation. For full details of the model, see \citet{Mutch2016}, and \citet{Qin2017} \citepalias[hereafter][]{Qin2017} for the black hole growth prescription.

This work introduces several extensions that have been made to \meraxesns: i) we now split galaxies into bulge and disc components, instead of the original disc-only model, ii) we introduce a new mechanism for black hole growth, iii) we implement a new model for determining galaxy disc sizes, iv) we use a more physical star formation prescription, and v) the angular momenta of the gas and stellar discs are now tracked throughout the evolution of the galaxy. These changes are discussed in detail in the following sections.

\subsection{Bulge formation and growth}
\label{sec:BulgeGrowth}
Previously published versions of \meraxes made no attempt to decompose a galaxy into its morphological components, instead assuming that all cold gas and stars in a galaxy are contained in a disc component. Following previous semi-analytic models \citep[e.g.][]{Croton2006a,Lucia2006,Guo2011,Menci2014,Tonini2016}, we expand the model to include a second galaxy component---galaxy bulges. These are assumed to contain no gas, and have no angular momentum, for simplicity. Our model for bulge growth is analogous to that of \citet{Tonini2016}, with bulges growing during galaxy mergers and disc instabilities. As in \citet{Tonini2016}, we split galaxy bulges into two types---merger-driven and instability-driven. We assume that the instability-driven bulge is disc-like, with a mass distribution that is flattened in the disc direction, whereas the merger-driven bulge is spheroidal.
These bulges are somewhat comparable to observed classical and `pseudo' bulges---which are typically built by mergers and disc instabilities, respectively \citep[see e.g.][]{Kormendy2004,Gadotti2009}---although they are classified by their observed properties and not their formation mechanism.

For a diagrammatic depiction of the bulge growth model, see figure 1 of \citet{Tonini2016}.

\paragraph*{Galaxy mergers:} 
When two galaxies merge, we first assume that the gas discs of the primary and secondary galaxies add, using the disc addition process described in Section \ref{sec:RadiusCalculation}. 

Galaxy mergers can induce an efficient burst of star formation, by causing strong shocks and turbulence which can drive cold gas towards the inner regions of the parent galaxy. 
We assume that in a merger with merger ratio $\gamma=M_{\textrm{2}}/M_{\textrm{1}} \geq 0.01$ (where $M$ is the total galaxy mass and the subscripts 1 and 2 herein denote the primary and secondary galaxy, respectively) the remnant undergoes a merger-driven starburst. This causes mass $M_{\textrm{burst}}=0.57 \gamma^{\ 0.7} M_{\textrm{cold,1}}$ to be converted from cold gas into stars \citep{Somerville2001,Mutch2016}, where $M_{\textrm{cold}}$ is the mass of cold gas in the galaxy. Any gas that is not converted into stars during this burst remains in a gas disc, regardless of merger ratio.  

To determine where these new stars are added, we consider the morphology of the primary galaxy, as we assume that the dominant dynamical component of the primary galaxy will regulate where the mass from the satellite will be deposited \citep[see][]{Tonini2016}. 
If the primary is dominated by a discy component---either the stellar disc, if $M_{\ast\textrm{ disc,1}}/M_{\ast\textrm{,1}}>0.5$, or the instability-driven bulge, if $M_{\ast\textrm{ disc,1}}/M_{\ast\textrm{,1}}\leq0.5$ and $M_{\textrm{MDB,1}}/M_{\ast\textrm{,1}}<0.5$ (where $M_{\ast\textrm{ disc}}$ is the mass of the stellar disc, $M_{\ast}$ is the total stellar mass of the galaxy, and $M_{\textrm{MDB}}$ is the mass of the merger-driven bulge)---the mass deposition is likely to occur in the plane of the disc, and so these stars are added to the instability-driven bulge. 
Otherwise, the dominant mass component of the primary is the spheroidal merger-driven bulge, and so we assume that the newly formed stars will accumulate in shells around it; stars from the burst are added to the merger-driven bulge. 

After the starburst, the black holes of the primary and secondary are combined (we term this `BH--BH coalescence'). 
The remnant undergoes merger-driven quasar-mode black hole growth given by
\begin{equation}
\Delta M_{\textrm{BH,}m} = \frac{k_m M_{\textrm{cold}}}{1+(280\textrm{km s}^{-1}/V_{\textrm{vir}})^2}
\label{eq:BHM}
\end{equation}
where $V_{\textrm{vir}}$ is the virial velocity of the host dark matter halo, and efficiency $k_m=k_c \gamma$ with $\gamma$ the merger ratio, and $k_c$ a free parameter in our model  \citepalias{Qin2017}.

Finally, we consider the addition of the stellar component of the secondary galaxy to the primary.
If a galaxy undergoes a major merger, which we define as a merger with $\gamma>0.1$\footnote{In \citet{Mutch2016} and \citetalias{Qin2017} we define a major merger as one with $\gamma>0.3$. Here we modify the definition to $\gamma>0.1$, to produce better agreement between our model and the observed bulge fraction of high-mass galaxies---see Section \ref{sec:Calibration}}, all stars and metals from both galaxies are placed into the merger-driven bulge of the remnant---major mergers are assumed to form pure bulge galaxies.
In a minor merger ($\gamma \leq 0.1$) where the primary is disc-dominated ($M_{\ast\textrm{ disc,1}}/M_{\ast\textrm{,1}}>0.5$), we assume that the stars from the secondary are added to the primary's disc. This causes a gravitational instability such that  $M_{\ast\textrm{,2}}$ is taken from the disc of the primary and placed into its instability-driven bulge \citep[see][]{Tonini2016}.
In a minor merger where the primary is bulge-dominated ($M_{\ast\textrm{ disc,1}}/M_{\ast\textrm{,1}}\leq0.5$), the secondary's mass is simply added to the primary's bulge (either the instability-driven bulge if $M_{\rm{MDB,1}}/M_{\ast\rm{,1}}<0.5$, or the merger-driven bulge otherwise).
 
\paragraph*{Disc instabilities:} For thin galaxy discs with an exponential surface density (with scale-radius $R_{\textrm{s}}$) and flat rotation curve (with velocity $V_{\textrm{disc}}$), a disc is stable if its mass $M_{\textrm{disc}}<V_{\textrm{disc}}^2R_{\textrm{s}}/G=M_{\textrm{crit}}$ \citep{Efstathiou1982,Mo1998}.
Thus, we assume that if the galaxy disc accretes enough material such that $M_{\textrm{disc}}>M_{\textrm{crit}}$, then it is no longer in dynamical equilibrium and so an amount $M_{\textrm{unstable}}=M_{\textrm{disc}}-M_{\textrm{crit}}$ of this disc mass is unstable. Here, we take $M_{\textrm{disc}}$ as the combined mass of both gas and stars, and $R_{\textrm{s}}$ as the mass-weighted scale radius of the stellar and gas discs. We assume that the ratio of unstable stars to gas is equal to their total mass ratio, so $M_{\textrm{unstable, }\ast}=M_{\textrm{unstable}} M_{\ast\textrm{ disc}}/M_{\textrm{disc}}$.
To return the disc to equilibrium, we assume that $M_{\textrm{unstable, }\ast}$ of stars is transferred from the disc to the instability-driven bulge, with their current metallicity. 

We then assume that the disc instability can drive black hole growth \citep[as seen in e.g.][]{Fanidakis2011,Hirschmann2012,Menci2014,Croton2016,Irodotou2018}.
Following the prescription of \citet{Croton2016}, when the galaxy undergoes a disc instability, its black hole mass grows by
\begin{equation}
\Delta M_{\textrm{BH,}i} = \frac{k_i M_{\textrm{unstable gas}}}{1+(280\textrm{km s}^{-1}/V_{\textrm{vir}})^2},
\label{eq:BHI}
\end{equation}
where $k_i$ is a free parameter which modulates the strength of the black hole accretion. This growth mechanism is analogous to the merger-driven quasar mode that we implement, which is described by Equation \ref{eq:BHM}. Note that while \citet{Croton2016} choose to have $k_i=k_c$, we consider these two separate free parameters as it produces a better agreement with the local black hole--bulge mass relation. Physically, we know of no expectation for the two growth modes to have the same black hole growth efficiency due to the different physical processes that drive them.
After tuning our model to observed galaxy stellar mass functions at $z=8$--0 and the black hole--bulge mass relation at $z=0$ (see Section \ref{Verification}) we find values of these parameters of $k_c=0.03$ and $k_i=0.02$: quasar-mode black hole growth due to galaxy mergers is 1.5 times more efficient at growing black holes than instability-driven growth in our model. 

Any excess gas will migrate towards the denser galaxy centre and form stars.
Hence in our model, after this black hole growth, the remainder of the unstable gas is consumed in a 100 per cent efficient starburst, with the stars formed added to the instability-driven bulge.

\subsection{Density-dependent star formation}
\label{sec:SF}
We implement a new density-dependent star formation prescription equivalent to that in \citet{Tonini2016}.
This uses the mass of cold gas above the critical density for star formation to determine how many stars are formed.
To calculate this critical mass from the critical surface density, the original \meraxes star formation prescription, used commonly in other semi-analytic models such as \citet{Croton2006a,Croton2016}, assumed that the mass in the gas disc was evenly distributed out to a radius of $3R_{\textrm{s,g}}$, where $R_{\textrm{s,g}}$ is the scale-length of the gas disc. Models such as \citet{Lucia2008}, \citet{Lagos2011} and \citet{Tonini2016} improve this by assuming that the gas in the disc is exponentially distributed. This will result in a lower fraction of gas being above the critical density over longer periods of time, reducing the burstiness of the star formation.

The gas surface density threshold for star formation is $\Sigma_{\textrm{crit}}=10 M_\odot/\textrm{pc}^2$ (Kormendy \& Kennicutt 2004).  For gas discs with an exponential surface density profile  $\Sigma(r)=\Sigma_0\exp\left(-r/R_{\textrm{s,g}}\right);\ \Sigma_0=M_{\textrm{cold}}/(2\pi R_{\textrm{s,g}}^2)$, where the scale radii $R_{\textrm{s,g}}$ are determined as outlined in Section \ref{sec:RadiusCalculation}, the radius at which the surface density drops below this value is $r_{\textrm{crit}}=R_{\textrm{s,g}}\ln(\Sigma_0/\Sigma_{\textrm{crit}}).$
The total mass of gas inside this radius is
\begin{equation}
M_{\textrm{crit}}=M_{\textrm{cold}}\left(1-\exp{\left(\frac{-r_{\textrm{crit}}}{R_{\textrm{s,g}}}\right)}\left(1+\frac{r_{\textrm{crit}}}{R_{\textrm{s,g}}}\right)\right)
\end{equation}
and so we assume that this amount of gas is capable of forming stars since it is above the critical density threshold for star formation. New stars form at a rate of $\textrm{SFR}=\epsilon M_{\textrm{crit}}/t_{\textrm{dyn}}=\epsilon M_{\textrm{crit}}V_{\textrm{vir}}/r_{\textrm{crit}}$, where $t_{\textrm{dyn}}=r_{\textrm{crit}}/V_{\textrm{vir}}$ is the dynamical time of the disc and $\epsilon$ is a free parameter describing the efficiency of star formation. Therefore, over the snapshot time length of $dt$, the total mass of new stars which have formed is
\begin{equation}
\delta M_\ast=\epsilon M_{\textrm{crit}} V_{\textrm{vir}} /r_{\textrm{crit}} ~dt.
\end{equation}

\subsection{Tracking galaxy radius and angular momentum}
\label{sec:RadiusCalculation}
The simulated properties of dark matter haloes can jump significantly between adjacent snapshots due to errors in the halo-finding process \citep[e.g. central-satellite switching, ejected cores, and the `small halo problem'; for details see][]{Poole2017}. In previous versions of \meraxes \citep[e.g.][]{Mutch2016,Liu2016}, the scale radius of the disc $R_s$ is approximated from the radius $R_{\textrm{vir}}$ and spin $\lambda=|\textbf{j}_H|/(\sqrt{2}R_{\textrm{vir}}V_{\textrm{vir}})$ of the host halo---$R_s=R_{\textrm{vir}}\lambda/\sqrt{2}$---and it is assumed that the rotational velocity of the disc is equal to the maximal circular velocity of the halo. This means that the galaxy properties can also jump significantly between adjacent snapshots due to such errors. Clearly this is unphysical, and decoupling the galaxy properties from the instantaneous halo properties to remove these errors would be ideal. We therefore introduce a new method for determining the scale radius of galaxy discs, in which the disc is evolved incrementally in response to changes that have occurred at each snapshot. 
In order to do so, we consider any changes to the mass of the disc as a sum (or subtraction) of two discs: the existing disc and a pseudo-disc of material to be added (or removed), and determine the size of the resulting disc using conservation of energy and angular momentum arguments. This requires that we also track the vector angular momentum of both the stellar and cold gas discs. Note that we do not attempt to track the angular momentum of the bulge, and simply assume it to be zero. The method of calculating disc radii and tracking angular momentum is as follows.

We assume that the stellar and gas discs have:
\begin{enumerate}[i.]
\item Constant velocity profile $V(r)=V$ (i.e. the rotation curve of the disc is flat)
\item An exponential surface density profile, $\Sigma(r)=(M_{\textrm{disc}}/2\pi R_s^2)\exp{(-r/R_s)}$, as in \citet{Mo1998}
\end{enumerate}  
The total energy of such a disc is 
\begin{dmath}
E_{\textrm{tot}}=-\frac{GM_{\textrm{disc}}}{R_s^2} \int_0^\infty M(<r)\textrm{e}^{(-r/R_s)}dr+
\frac{1}{2}M_{\textrm{disc}} V^2.
\end{dmath}
For a stellar disc in a galaxy with a bulge, gas disc and surrounding dark matter halo, 
$M(<r)=M_{\ast\textrm{,disc}}(<r)+M_{\textrm{cold}}(<r)+M_{\textrm{halo}}(<r)+M_{\textrm{bulge}}(<r)$.
For discs with exponential surface density profiles, $M_{\textrm{disc}}(<r)=M_{\textrm{tot}}(1-(1+r/R_s)\exp(-r/R_s))$. Assuming that both the halo and the bulge are singular isothermal spheres such that $M_{\textrm{bulge}}(<r)=V_{\textrm{bulge}}^2r/G$ and $M_{\textrm{halo}}(<r)=V_{\textrm{vir}}^2r/G$, we have:
\begin{dmath}
E_{\textrm{tot}}=-\frac{GM_{\ast\textrm{,disc}}}{R_{s,\ast}}\left(\frac{M_{\ast\textrm{,disc}}}{4}+M_{\textrm{cold}}\frac{R_{s,\ast}^2}{(R_{s,\ast}+R_{s,g})^2}\right)-M_{\ast\textrm{,disc}}V_{\textrm{vir}}^2-M_{\ast\textrm{,disc}}V_{\textrm{bulge}}^2+\frac{1}{2}M_{\ast\textrm{,disc}} V^2,
\end{dmath}
where $R_{s,\ast}$ and $R_{s,g}$ are the scale radii of the stellar and gas discs, respectively.

The potential energy from the gas disc and the stellar disc itself are negligible relative to the potential energy from the halo and bulge: neglecting these gives the simplified total energy equation
\begin{equation}
E_{\textrm{tot}}\simeq-M_{\ast\textrm{,disc}}V_{\textrm{bulge}}^2-M_{\ast\textrm{,disc}}V_{\textrm{vir}}^2+\frac{1}{2}M_{\ast\textrm{,disc}} V^2.
\end{equation}
An identical argument applies to the energy of a gas disc.

Now consider the sum (or subtraction) of two such discs, the existing disc and a pseudo-disc of material to be added (or removed). Both discs are assumed to reside at the centre of the same halo, and are associated with the same bulge component. If these discs have masses $M_1$ and $M_2$, scale radii $R_1$ and $R_2$, and velocities $V_1$ and $V_2$, then from conservation of energy, $E_1+E_2=E_{\textrm{new}}$, which leads to, upon cancellation of the potential terms:
\begin{equation}
\frac{1}{2}M_1 V_1^2+\frac{1}{2}M_2 V_2^2=\frac{1}{2}(M_1+M_2) V_{\textrm{new}}^2.
\label{EEqn}
\end{equation}

Assuming conservation of angular momentum, $\textbf{J}_1+\textbf{J}_2=\textbf{J}_{\textrm{new}}$, the angular momentum of the combined disc $\textbf{J}_{\textrm{new}}$ is easily calculated by summing the individual angular momentum vectors of each disc, which are tracked by the model. Under the assumptions \textit{i} and \textit{ii}, the angular momentum is
\begin{align}
|\textbf{J}|&=\int_0^\infty \Sigma (r)2\pi r^2 V dr \nonumber\\ 
&=\frac{M_{\textrm{disc}}V}{R_s^2} \int_0^\infty \textrm{e}^{(-r/R_s)} r^2 dr \nonumber\\
&=2M_{\textrm{disc}}VR_s.
\label{AngMomEqn}
\end{align} 
Using conservation of energy and angular momentum (Equations \ref{EEqn} and \ref{AngMomEqn}), the new disc scale length can be calculated using the galaxy's existing properties and the properties of the new material added:

\begin{numcases}{}
V_{\textrm{new}}=\sqrt{\frac{M_1V_1^2+M_2V_2^2}{M_1+M_2}}\\
R_{\textrm{new}}=\frac{|\textbf{J}_{\textrm{new}}|}{2V_{\textrm{new}}(M_1+M_2)}
\end{numcases}

This method is applied when the galaxy undergoes several different processes, as follows.
\paragraph*{Gas cooling:} 
In each dark matter halo we assume that there is a hot reservoir of pristine primordial gas, with baryon fraction $f_b=\Omega_b/\Omega_m$. Some fraction of this gas will then cool and condense into a gas disc. We assume that the hot halo gas is settled into the dark matter halo such that $V_{\textrm{hot}}=V_{\textrm{vir}}$ and $\textbf{j}_{\textrm{hot}}=\textbf{j}_H$. Then, we assume that when this gas cools on to the galaxy, it forms in a disc with velocity and specific angular momentum $V_{\textrm{cold}}=V_{\textrm{hot}}=V_{\textrm{vir}}$ and $\textbf{j}_{\textrm{cold}}=\textbf{j}_{\textrm{hot}}=\textbf{j}_{H}$, and thus with scale radius $R=|\textbf{j}_{H}|/(2V_{\textrm{vir}})$. If the galaxy already contains some gas, this new gas is added to the existing gas disc.

\paragraph*{Star formation:} 
If the gas disc exceeds some critical density (see Section \ref{sec:SF}), stars will form. Since these stars are originating in the disc, they are assumed to form in a stellar disc with $R=R_{\textrm{cold}}$, $V=V_{\textrm{cold}}$ and $\textbf{j}=\textbf{j}_{\textrm{cold}}$. This disc of new stars is added to the existing stellar disc and subtracted from the existing gas disc (by taking the mass $M_2$ to be negative), and the total angular momentum of these new stars is transferred from the gas to the stellar disc.

Stars can also form in our model via bursts caused by disc instabilities or mergers. If such a burst occurs, our model prescribes that new stars, with total mass $\Delta M_\ast$, form in the bulge and not the stellar disc. In this case, the stellar disc is unaffected, while the gas disc loses some mass $\Delta M_{\textrm{cold}}=-\Delta M_\ast$. We assume that the new stars form with zero angular momentum, so that the angular momentum of the gas disc is conserved. Assuming that the velocity of the gas disc remains constant, we use 
\begin{equation}
\label{eq:radius_mod}
r_{\textrm{new}}=\frac{|\textbf{J}_{\textrm{cold}}|}{2V_{\textrm{cold}}(M_{\textrm{cold}}+\Delta M_{\textrm{cold}})}=\frac{r_{\textrm{original}}}{1-\Delta M_\ast/M_{\textrm{cold}}}
\end{equation} 
to update the radius (note that mass is removed from the disc, and so its radius will increase).

\paragraph*{Mergers:} In a minor merger, the stellar and gas discs of the secondary galaxy are added to those of the primary galaxy, using the disc addition process described above.

To determine the angular momentum of the merger remnant, the \citet{Tonini2016} prescription assumes that the gas from the secondary is stripped and settles into the dark matter halo before being accreted by the primary, retaining no `memory' of the disc structure. In such a scenario, the angular momentum of the stripped material will be $\textbf{J}_{\textrm{stripped}}=M_{\textrm{sat}}\times \textbf{j}_{\textrm{FOF group of sat}}$, and so the angular momentum of the merger remnant will be $\textbf{J}_{\textrm{new}}=\textbf{J}_{\textrm{1}}+\textbf{J}_{\textrm{stripped}}=\textbf{J}_{\textrm{1}}+M_{\textrm{sat}}\times \textbf{j}_{\textrm{FOF group of sat}}$.
\citet{Lee2018} find that the majority of satellite galaxies experience little change to their angular momentum after their infall into a cluster. In contrast to the \citet{Tonini2016} model, we therefore assume that the secondary's disc retains its angular momentum, instead of being stripped.
To determine the angular momentum of the merger remnant, we thus simply add the secondary's angular momentum, as tracked throughout its history, to that of the primary: $\textbf{J}_{\textrm{new}}=\textbf{J}_{\textrm{1}}+\textbf{J}_{\textrm{2}}$.

In a major merger, the stellar component of the remnant galaxy is assumed to become purely a bulge, and so the scale length, velocity and angular momentum of the stellar disc are set to zero. We assume that the gas disc remains intact, with the gas discs of both galaxies added together, where they form more stars (see Section \ref{sec:BulgeGrowth}).

\paragraph*{Supernovae:} 
Supernovae convert material from stars into gas. 
We assume that a factor of $(1-M_{\textrm{bulge}}/M_{\ast\textrm{,total}})$ of the stars going supernova are in the stellar disc. The proportion of stars in the disc and bulge going supernovae may be different as they may have stellar populations with different ages; however, we do not expect this to significantly change our galaxy sizes.

For supernovae in the galaxy disc, we assume that the fraction of stars going supernova relative to the number of stars at any radius is constant; the surface density of supernova and therefore of the gas disc that they form is proportional to the stellar disc surface density. 
Therefore, we assume that the new gas disc caused by supernovae has the same velocity, specific angular momentum and scale radius $R=|\textbf{j}_\ast|/(2V_\ast)$ as the stellar disc.
This disc of new gas is added to the existing gas disc, and subtracted from the existing stellar disc, with the angular momentum transferred from the stellar disc to the gas disc. 

We treat supernovae in the galaxy bulge differently, as we do not model bulge sizes or angular momenta, and therefore do not track the orbital parameters of the bulge stars going supernova or the gas that they eject. To account for this, we assume that the gas from supernovae in the bulge is ejected into the surrounding halo, where it will approach equilibrium with the surrounding dark matter. This gas will settle into a disc with the velocity and specific angular momentum of the host halo, with scale radius $R=|\textbf{j}_H|/(2V_{\textrm{vir}})$. 
This disc is added to the existing gas disc, as is the total angular momentum of this new gas.  
Our assumption for the ejected gas from the bulge is motivated by the expectation that the bulge would have some angular momentum. However, we note that a more straightforward choice would be to assume that the ejected gas has no angular momentum. We ran \meraxes with this alternative assumption and found no differences in our results, due to the minimal mass ejected by bulge supernovae.

Some cold disc gas is reheated by supernova feedback. This has a total angular momentum $\textbf{J}_{\textrm{reheat}}=M_{\textrm{reheat}}~\textbf{j}_{\textrm{cold}}$, which is removed from the gas disc. This angular momentum is effectively given to the hot gas, however we do not track the angular momentum of the hot halo as we assume it is in equilibrium with the dark matter halo. Extending our model to track the hot gas angular momentum, as in the more complex model of \citet{Hou2017}, for example, is left for future work.

\paragraph*{Disc instabilities:} When a disc is gravitationally unstable, we assume some stars from the disc are transferred to the bulge. In this scenario, instead of considering the subtraction of a pseudo-disc, we follow the \citet{Tonini2016} prescription. While the angular momentum of the disc is conserved, its mass decreases by $\Delta M_\ast$. Assuming that the velocity of the disc remains constant, we use
\begin{equation}
r_{\textrm{new}}=\frac{|\textbf{J}_{\ast}|}{2V_{\ast}(M_{\ast}-\Delta M_{\ast})}=\frac{r_{\textrm{original}}}{1-\Delta M_\ast/M_{\ast}}
\end{equation} 
to update the radius, which will increase as the mass decreases.

\paragraph*{AGN feedback:} Quasar mode AGN feedback reheats some of the gas in the disc. We assume that the same fraction of gas gets heated at all locations in the disc, so the scale length of the reheated gas disc is equal to that of the original disc ($R_1=R_2$), and the two discs have the same velocities ($V_1=V_2$). 
In reality, the central AGN would preferentially expel gas from the centre of the galaxy; our simplistic assumption could be improved in future work by following the approach of \citet[][]{Stevens2016}, for example.

Under our assumption, since $\textbf{j}=2VR$, we are effectively assuming that the specific angular momentum of the disc is conserved during this process. Therefore, we decrease the total angular momentum of the disc by $m_{\textrm{reheated}}\textbf{j}$. This disc of reheated gas is subtracted from the existing gas disc, with the mass given to the hot gas reservoir. 
We ensure that the scale length and velocity are zero if the mass of the disc is zero.
\\

Finally, note that in general, $\textbf{J}_1$, $\textbf{J}_2$ and $\textbf{J}_{\textrm{new}}$ will not be aligned.
We define the new orientation of the disc using the direction of $\textbf{J}_{\textrm{new}}$. Consider the projection of $\textbf{J}_1$ and $\textbf{J}_2$ on to this direction:
\begin{numcases}{}
\textbf{J}_{1p}=|\textbf{J}_1| \sin{\theta_1} \frac{\textbf{J}_{\textrm{new}}}{|\textbf{J}_{\textrm{new}}|}=2M_1V_1R_1\sin{\theta_1} \frac{\textbf{J}_{\textrm{new}}}{|\textbf{J}_{\textrm{new}}|}\\
\textbf{J}_{2p}=|\textbf{J}_2| \sin{\theta_2}\frac{\textbf{J}_{\textrm{new}}}{|\textbf{J}_{\textrm{new}}|}=2M_2V_2R_2\sin{\theta_2} \frac{\textbf{J}_{\textrm{new}}}{|\textbf{J}_{\textrm{new}}|}.
\end{numcases}
Now $\textbf{J}_{\textrm{new}}=\textbf{J}_1+\textbf{J}_2$, so $|\textbf{J}_{\textrm{new}}|=|\textbf{J}_{1p}|+|\textbf{J}_{2p}|$, that is,
\begin{equation}
|\textbf{J}_{\textrm{new}}|=2M_{\textrm{new}}V_{\textrm{new}}R_{\textrm{new}}=2M_1V_1R_1\sin{\theta_1}+2M_2V_2R_2\sin{\theta_2}.
\end{equation}
This can be interpreted as summing two discs that are projected on to the new plane, with projected radii $R_1\sin{\theta_1}$ and $R_2\sin{\theta_2}$. Therefore, our method is a simpler version of the
\citet{Stevens2016} method of projecting the two discs prior to their addition, which captures the effect of more compact discs.

\section{Model Verification}
\label{Verification}
In this section we detail how the free parameters in \meraxes are calibrated, and show that this calibrated model can reproduce a range of observations.
All analysis henceforth is conducted using model galaxies with $M_\ast>10^7M_\odot$, the resolution limit of \meraxes when run on the Tiamat simulation \citep[see][]{Mutch2016}. \meraxes includes a limited prescription for modelling the physics of satellite galaxies, assuming a maximally efficient hot halo stripping scenario and excluding potentially important dynamical effects which could alter satellite sizes or morphologies.  To ensure that our conclusions are robust, we therefore only consider galaxies classified as centrals, except in the stellar mass and luminosity functions, which require the full galaxy sample.  We leave a detailed exploration of the effect of including satellite galaxies to future work \citep[however see, for example,][]{Zoldan2018}.

Note that we use the same parameter values for both \textit{Tiamat} and \textit{Tiamat-125-HR}, and use both simulations to tune the model: \textit{Tiamat} for matching $z\geq2$ observations and \textit{Tiamat-125-HR} for $z<2$.
The stellar and black hole mass functions from the \textit{Tiamat} and \textit{Tiamat-125-HR} simulations at $z=2$ are converged above $M_\ast=10^{8.6}M_\odot$ and $M_{\textrm{BH}}=10^{7.1}M_\odot$, respectively. At $z=0$, we note that this convergence limit is $M_\ast=10^{8.65}M_\odot$ and $M_{\textrm{BH}}=10^{7.35}M_\odot$ (see Appendix \ref{Appendix:Resolution} for details). We therefore place no significance on the trends observed with \textit{Tiamat-125-HR} for lower stellar and black hole masses, as these may be affected by resolution effects. These convergence limits are shown on all relevant plots. 


\begin{figure*}
\begin{center}
\includegraphics[scale=0.9]{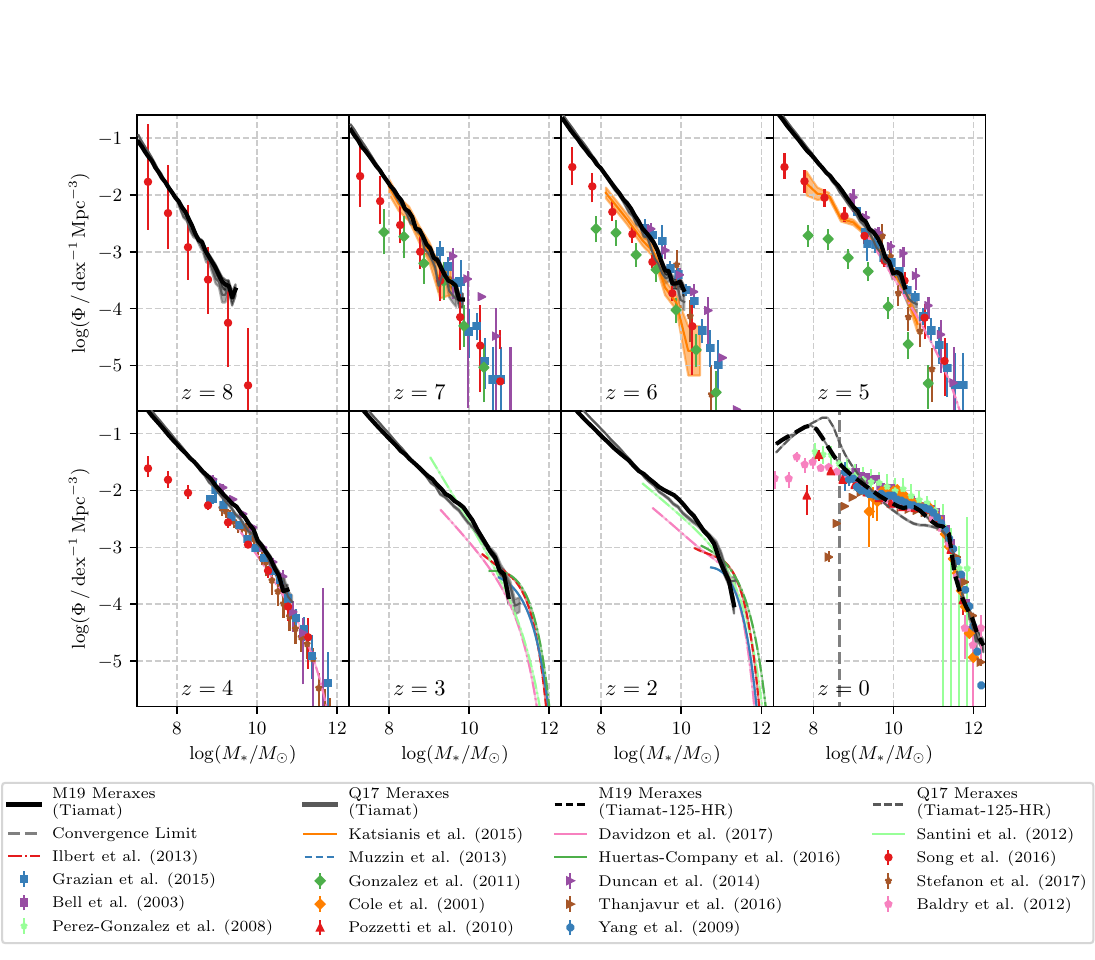}
\caption{Galaxy stellar mass functions at $z=8$--0 from the \citetalias{Qin2017} \meraxes (grey) and our new model (M19, black) applied to \textit{Tiamat} (solid) and \textit{Tiamat-125-HR} (dashed), compared to a range of observational data (see legend). \meraxes is calibrated such that these observed stellar mass functions are reproduced. The vertical grey dotted line indicates the stellar mass below which  \textit{Tiamat} and \textit{Tiamat-125-HR} are not converged, and thus where galaxies from \textit{Tiamat-125-HR} can be subject to resolution effects (see Appendix \ref{Appendix:Resolution}).}
\label{GSMF}
\end{center}
\end{figure*}

\subsection{Calibration}
\label{sec:Calibration}

\begin{table}
\centering
\caption{Optimal values of \meraxes parameters that are either introduced or discussed in this work, or have different values from those used in \citetalias{Qin2017} due to the model update and re-calibration we have implemented}
\label{tab:Parameters}
\begin{threeparttable}
\begin{tabular}{ccc}
\hline 
Parameter & \citetalias{Qin2017} & M19 \\ 
\hline 
Star formation efficiency \tnote{a}  & 0.08 & 0.03 \\ 
Supernova ejection efficiency \tnote{b} & 1 & 0.5\\
Minimum merger ratio for major merger & 0.3 & 0.1 \\ 
Minimum merger ratio for starburst & 0.1 & 0.01 \\ 
Black hole seed mass ($M_\odot$) & $10^3$ & $10^4$ \\ 
Merger-driven black hole growth efficiency ($k_c$) \tnote{c} & 0.05 & 0.03 \\ 
Instability-driven black hole growth efficiency ($k_i$) \tnote{d} & - & 0.02 \\ 
Radio mode black hole growth efficiency \tnote{e} & 0.3  & 0.003 \\ 
Black hole efficiency of converting mass to energy \tnote{f} & 0.06  & 0.2 \\ 
Opening angle of AGN radiation \tnote{g} & $80^{\circ} $ & $30^{\circ} $ \\ 
\hline 
\end{tabular} 
\begin{tablenotes}
\item[a] \citet{Mutch2016} equation 7 
\item[b] \citet{Mutch2016} equation 13. Note that in \citet{Mutch2016} this value was 0.5, as it is now.
\item[c] Equation \ref{eq:BHM}, \citetalias{Qin2017} equation 19
\item[d] Equation \ref{eq:BHI}
\item[e] \citetalias{Qin2017} equation 14
\item[f] \citetalias{Qin2017} equation 15
\item[g] \citetalias{Qin2017} section 3.3
\end{tablenotes}
\end{threeparttable}
\end{table}

\begin{figure*}
\begin{center}
\includegraphics[scale=0.9]{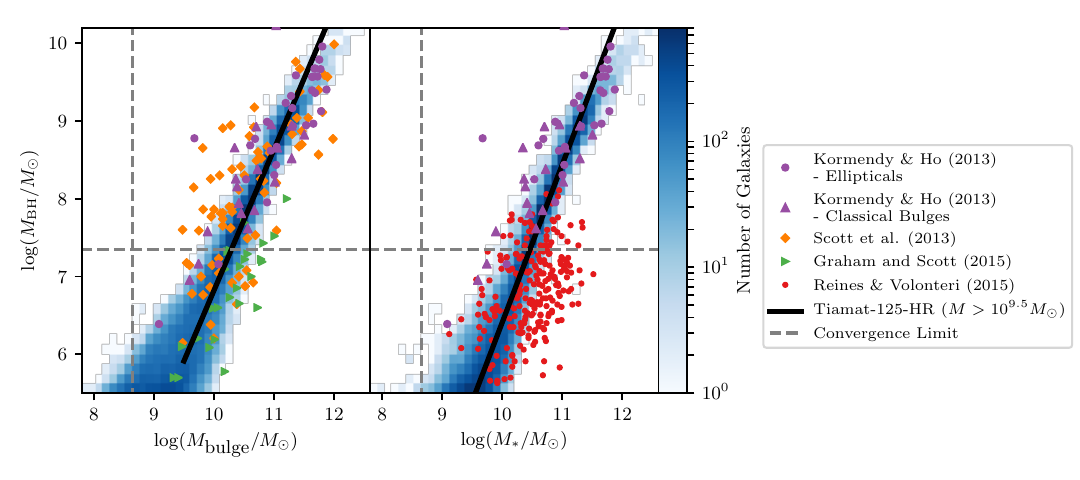}
\caption{\textit{Left panel:} The $z=0$ black hole--bulge mass relation, and \textit{Right panel:} the $z=0$ black hole--total stellar mass relation, for our model applied to \textit{Tiamat-125-HR} (blue density plot). Only galaxies classified as centrals are shown.
A range of observations are also plotted \citep[][see legend]{Kormendy2013,Scott2013,Graham2015,Reines2015}. A best-fitting line for our model galaxies with $M>10^{9.5} M_\odot$ is also shown (solid line).
\meraxes is calibrated to fit the canonical \citet{Kormendy2013} observations in the high stellar and bulge mass regime. 
The grey dotted lines indicate the stellar and black hole masses below which  \textit{Tiamat} and \textit{Tiamat-125-HR} are not converged, and thus where galaxy properties from \textit{Tiamat-125-HR} are more uncertain (see Appendix \ref{Appendix:Resolution}). }
\label{Magorrianz0}
\end{center}
\end{figure*}

The free parameters in \meraxes are tuned to match the observed stellar mass functions at $z=8$--0 (Figure \ref{GSMF}) and the $M_{\textrm{BH}}$--$M_{\textrm{bulge}}$ relation at $z=0$ (Figure \ref{Magorrianz0}).
We calibrate to the stellar mass function at a range of redshifts, as this has been shown to provide a tight constraint on both the star formation efficiency and supernova feedback parameters \citep[see e.g.][]{Mutch2013,Henriques2013}.
We use the $M_{\textrm{BH}}$--$M_{\textrm{bulge}}$ relation to calibrate the black hole growth parameters; note that we use the $M_{\textrm{BH}}$--$M_{\textrm{bulge}}$ relation instead of the black hole mass function as in \citetalias{Qin2017}, as the $M_{\textrm{BH}}$--$M_{\textrm{bulge}}$ relation is a more direct observable than the black hole mass function, and our model now has the capability of modelling bulge masses. We place an emphasis on matching the canonical \citet{Kormendy2013} observations at high stellar and bulge masses, as this is a widely-used and reliable sample. We note that our model does not reproduce the wide scatter observed by \citet{Scott2013}, nor the \citet{Reines2015} population of high stellar mass, low black hole mass galaxies, albeit this sample falls under our resolution limits (see Figure \ref{Magorrianz0}).

To produce a median bulge fraction that is in better agreement with local observations at the highest stellar masses (see Section \ref{sec:lowzmorphology} and Figure \ref{discFraction}), we also modify the definition of a major merger to be one with $\gamma>0.1$, instead of $\gamma>0.3$ as used in \citet{Mutch2016} and \citetalias{Qin2017}\footnote{Semi-analytic models use a range of merger thresholds, such as 0.1 \citep{Henriques2015}, 0.2 \citep{IzquierdoVillalba2019}, and 1/3 \citep{Guo2011}, and this is often tuned to better reproduce morphologies \citep[e.g.][]{Henriques2015,IzquierdoVillalba2019}.}.

The parameter values discussed in this paper and those that differ from the \citetalias{Qin2017} values are shown in Table \ref{tab:Parameters}. 
The only parameters we have introduced with these model updates are the instability-driven black hole growth efficiency, and the efficiency at which mass is converted to luminosity upon accretion by a black hole, $\eta$:
$L_{\textrm{BH}} = \eta \dot M_{\textrm{BH}}c^2$,
where $c$ is the speed of light. In previous versions of \meraxes, this value was fixed at 0.06 \citep{Qin2017}. However, we add this free parameter to better model the high-mass end of both the stellar and black hole mass functions, and set it to $\eta=0.2$ through the parameter tuning \citep[see][for a discussion]{Croton2016}.

\subsection{Further validation}
\label{FurtherValidation}
Once the model has been tuned to reproduce the galaxy stellar mass functions and local black hole--bulge relations, we verify the changes to our model by comparing the output to a range of local observations: the stellar mass--disc size relation, bulge fractions, and angular momentum--mass relations.
Additionally, comparison of the high-redshift galaxy UV luminosity functions and the stellar mass--star formation rate relation can be seen in Appendix \ref{Appendix:Validation}.

\begin{figure}
\begin{center}
\includegraphics[scale=0.9]{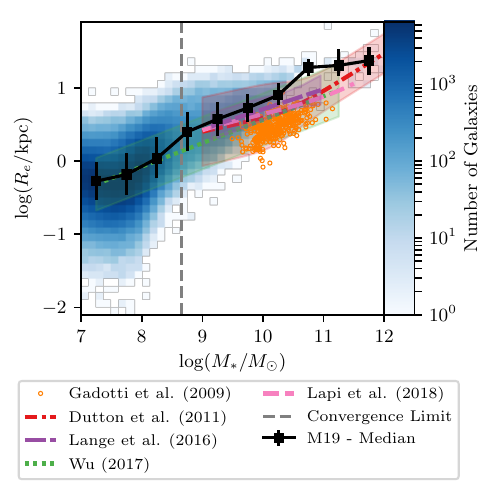}
\caption{The relation between the disc effective radius ($R_{e}$) and stellar mass ($M_\ast$) at $z=0$ for our updated \meraxes (M19, blue density plot). Only galaxies classified as centrals are shown. A range of observations are also plotted \citep[][see legend]{Gadotti2009,Dutton2010,Lange2016,Wu2017,Lapi2018}. The black squares and errorbars represent the median and 16th and 84th percentile ranges of galaxy disc sizes from our model, in mass bins with width 0.5 dex. The filled regions around the observations represent the 1$\sigma$ scatter for \citet{Dutton2010} and \citet{Lange2016}, and 2$\sigma$ for \citet{Wu2017}. The grey dotted line indicates the stellar mass below which  \textit{Tiamat} and \textit{Tiamat-125-HR} are not converged, and thus where galaxy properties from \textit{Tiamat-125-HR} are more uncertain (see Appendix \ref{Appendix:Resolution}).}
\label{discSize}
\end{center}
\end{figure}

\subsubsection{Low-redshift disc size--mass relation}

We present the sizes of model galaxies using the physical effective radius, or half-light radius, $R_e$. We estimate this as $R_e=1.678 R_s$, which is the half-light radius for a disc with an exponential surface density profile and a constant mass-to-light ratio.

Figure \ref{discSize} shows the relation between the stellar disc effective radius and stellar mass at $z = 0$ for \meraxes and a range of observations. 
The model agrees reasonably well with the \citet{Dutton2010}, \citet{Lange2016} and \citet{Lapi2018} observations, with the median matching their relations closely for $10^9<M_\ast/M_\odot<10^{10.5}$. At higher masses, our model median jumps to higher radii, though is still reasonably consistent with the observations; this jump is most likely a result of the lower sample size of high-mass galaxies, due to the limited simulation box size.
The observations of \citet{Wu2017} extend to much lower masses; the median of our model is highly consistent with this sample at $M_\ast<10^{8.5}M_\odot$, while being slightly higher yet still consistent with the observations at higher masses.

We note that \citet{Dutton2010}, \citet{Wu2017}, and \citet{Lapi2018} consider only disc galaxies, while the \citet{Lange2016} and \citet{Gadotti2009} samples have no clear morphology biases. 
These observations are all selected in the optical/near-infrared, except for the \citet{Wu2017} observations which are HI selected, and are generally volume-limited. \citet{Gadotti2009} note that their sample, however, contains a larger fraction of massive and more concentrated galaxies than a volume-limited sample. In addition, they use circular apertures, as opposed to more accurate elliptical apertures, leading to an underestimation of the sizes of galaxies. Indeed, Figure \ref{discSize} shows that \citet{Gadotti2009} observe smaller disc sizes relative to the other observations and our model for galaxies in $10^{9.5}<M_\ast/M_\odot<10^{11}$.

We conclude that our model is reproducing reasonable $z=0$ galaxy sizes, given the dispersion in available observational data.

\begin{figure*}
\begin{center}
\includegraphics[scale=1]{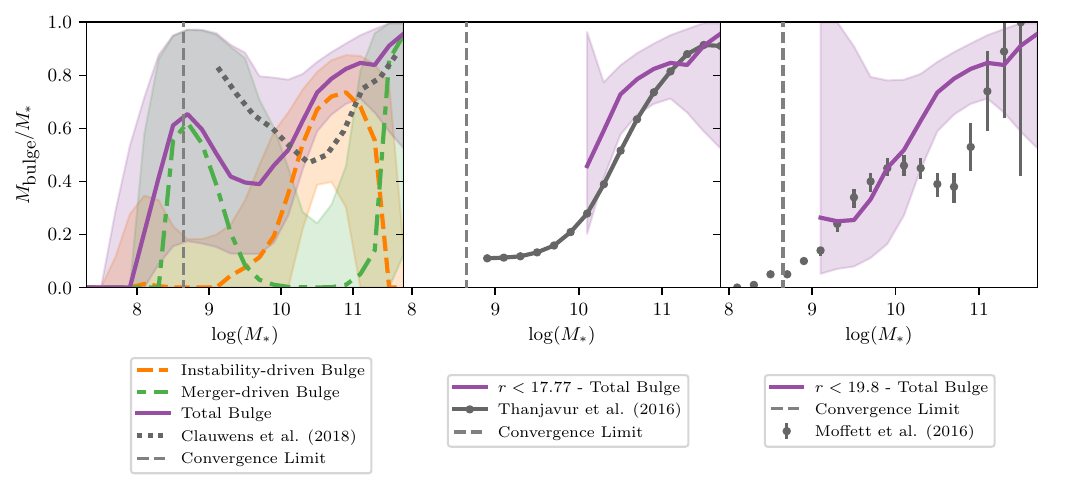}
\caption{The fraction of stellar mass contained in the bulge components of galaxies as a function of total stellar mass at $z=0.1$. Only model galaxies classified as centrals are shown. 
\textit{Left panel:} Purple solid lines show the median galaxy bulge fraction for the entire sample of central galaxies, in mass bins of width 0.2 dex, and the purple shaded regions show the 16th and 84th percentile range. Also shown are the median mass fractions of the instability-driven bulge (orange dashed) and merger-driven bulge (green dot-dashed), alongside their 16th and 84th percentile ranges. For comparison, the predictions of the EAGLE simulation at $z=0$ \citep{Clauwens2018} are also shown (grey dotted). 
\textit{Middle panel:} Purple solid lines show the median galaxy bulge fraction for the central galaxies with magnitude $r<17.77$, in mass bins of width 0.2 dex, and the purple shaded regions show the 16th and 84th percentile range.
Also shown are the observed $z=0.1$ SDSS bulge fractions from \citet{Thanjavur2016} binned by stellar mass (grey points and line); these observations have a magnitude limit of $r<17.77$.
\textit{Right panel:} Purple solid lines show the median galaxy bulge fraction for the central galaxies with magnitude $r<19.8$, in mass bins of width 0.2 dex, and the purple shaded regions show the 16th and 84th percentile range.
Also shown are the observed $0.002<z<0.06$ GAMA bulge fractions from \citet{Moffett2016}, binned by stellar mass (grey points and errors); these observations have a magnitude limit of $r<19.8$. 
In all panels, the grey dotted line indicates the stellar mass below which  \textit{Tiamat} and \textit{Tiamat-125-HR} are not converged (see Appendix \ref{Appendix:Resolution}).}
\label{discFraction}
\end{center}
\end{figure*}

\subsubsection{Galaxy morphologies at low redshift}
\label{sec:lowzmorphology}
To verify that our bulge model is producing galaxies with reasonable morphologies, we consider the median fraction of stellar mass contained within galaxy bulges as a function of total stellar mass at $z=0.1$ (Figure \ref{discFraction}). Note that we chose this redshift instead of $z=0$ as it is the median redshift of the \citet{Thanjavur2016} observations to which we compare our model, and this allows us to implement a magnitude cut to mimic their selection effects; however, the equivalent plots at $z=0$ are qualitatively identical.

Our model predicts that the majority of stellar mass for low mass galaxies ($M_\ast\simeq10^9M_\odot$) is contained in the bulge component, discs contribute a peak of $\sim60$ per cent of the stellar mass at $M_\ast\simeq10^{9.5}M_\odot$, and bulges again dominate at larger masses. This dip in the bulge-to-total mass ratio B/T is due to the masses at which the two bulge growth mechanisms are dominant; the merger-driven growth mode peaks for galaxies with $M_\ast\simeq10^9M_\odot$, while the instability-driven growth mode peaks at $M_\ast\simeq10^{11}M_\odot$. This suggests that high-mass galaxy discs are likely to be unstable and thus form bulges, while it is predominantly only less massive galaxies which form their bulges via mergers. We also see a steep increase in the merger-driven bulge fraction for the most massive galaxies, with major mergers forming giant ellipticals.  We note that at masses $M_\ast<10^{8.65}M_\odot$, where the \textit{Tiamat} and \textit{Tiamat-125-HR} simulations are not converged, the bulge fraction drops rapidly---this is most likely a resolution effect.

We compare our results to the SDSS observations of \citet{Thanjavur2016}, which are also shown in Figure \ref{discFraction}. 
We implement a magnitude cut of $r<17.77$, to match those observations; this magnitude cut restricts our sample to galaxies with $M_\ast>10^{10}M_\odot$.
For stellar masses of $\gtrsim10^{11} M_\odot$ the model predicts a bulge fraction in remarkable agreement with the observations, increasing with mass to $\simeq0.9$ at $M_\ast\simeq10^{11.5}M_\odot$.  For masses $10^{10}<M_\ast<10^{11}M_\odot$, our model overpredicts the total bulge fraction by approximately 10 per cent, though is consistent within the 16th--84th percentile range of simulated galaxy bulge fractions. 

We perform an equivalent comparison with the GAMA observations of \citet{Moffett2016} (Figure \ref{discFraction}). 
We implement a magnitude cut of $r<19.8$, to match those observations; this magnitude cut restricts our sample to galaxies with $M_\ast>10^{9}M_\odot$.
For stellar masses of $M_\ast\gtrsim10^{11} M_\odot$ and $M_\ast\lesssim10^{10} M_\odot$ the model predicts a bulge fraction in good agreement with the observations.  For masses $10^{10}<M_\ast/M_\odot<10^{11}$, our model overpredicts the total bulge fraction by up to 40 per cent. However, we note that their analysis does not include pseudobulges---this mass range is precisely where we predict the instability-driven or `pseudo' bulge to dominate, which may explain this discrepancy.

The SDSS and GAMA observations provide the most comprehensive and quantitative comparison, however, other studies have also investigated the bulge fraction as a function of stellar mass. \citet{Bundy2005} predict a similar trend, finding that E/S0 galaxies dominate at higher masses and spirals at lower masses, with the transition at (2--3)$\times10^{10}M_\odot$ at $z\simeq0.3$.
The trend our model predicts for the B/T ratio is qualitatively consistent with observations such as \citet{Conselice2006}, who found that the spiral fraction is largest in galaxies with $10^9<M_\ast<10^{11}M_\odot$, with ellipticals dominant at higher masses and irregular galaxies at lower masses. 
\citet{Simons2015} find that for $M_\ast>10^{9.5}M_\odot$, galaxies are mostly rotation dominated and disc-like, while below this mass galaxies can be either rotation-dominated discs or asymmetric or compact galaxies; however, we note that they only consider `blue' galaxies, so this comparison is limited. \citet{Wheeler2017} found that the majority of dwarf galaxies with $M_\ast<10^8M_\odot$ are dispersion-supported (i.e. bulgy; due to our resolution limit we cannot make a comparison with this study).

Our predictions are in reasonable agreement with those of the \citet{IzquierdoVillalba2019} semi-analytic model, who find that pseudo-bulges, or those caused by disc instabilities, are typically present in more massive galaxies, peaking at $M_\ast\simeq10^{10}M_\odot$, while ellipticals dominate at the highest masses due to major mergers turning the galaxies into pure bulges.
The predictions of our model are also qualitatively very similar to those of the EAGLE simulation \citep{Clauwens2018}, whose median B/T ratios as a function of total stellar mass at $z=0$ are overplotted in Figure \ref{discFraction}. \citet{Clauwens2018} see three phases of galaxy growth, with $M_\ast<10^{9.5}M_\odot$ galaxies mostly spheroidal, disorganised irregulars, $10^{9.5}<M_\ast<10^{11}M_\odot$ galaxies dominated by discs, and galaxies with $M_\ast>10^{11}M_\odot$ mostly elliptical \citep{Clauwens2018}. While the dip in their B/T is slightly deeper and occurs at $\sim0.5$ dex higher mass, the predictions are qualitatively similar to those of our model. 
We can adjust our galaxy morphologies by introducing an additional free parameter in the disc-instability criterion, $\eta$, such that $M_{\textrm{crit}}=\eta V_{\textrm{disc}}^2R_{\textrm{s}}/G$ \citep{Efstathiou1982}; here $\eta$ is a parameter that determines the importance of the self-gravity of the disc. We find that increasing $\eta$ from its default value of 1 deepens the dip in B/T and shifts it to higher masses, bringing our model to closer agreement with \citet{Clauwens2018}. However, we opt not to introduce this as an additional free parameter in the model \citep[as in e.g.][]{Lagos2018,IzquierdoVillalba2019}.
Our results are also consistent with the morphologies predicted by the Illustris hydrodynamical simulation \citep{Snyder2015}, which find at $M_\ast\simeq10^{10}M_\odot$ that galaxies are a combination of spirals and composite systems, shifting to ellipticals at higher masses. 

Our bulge model therefore produces reasonable galaxy morphologies, and bulge and disc masses, given the stellar mass functions and bulge fractions are in agreement with the observations.

\subsubsection{Angular momentum--mass relations}
\label{sec:lowzj}
 
\begin{figure}
\begin{center}
\includegraphics[scale=0.9]{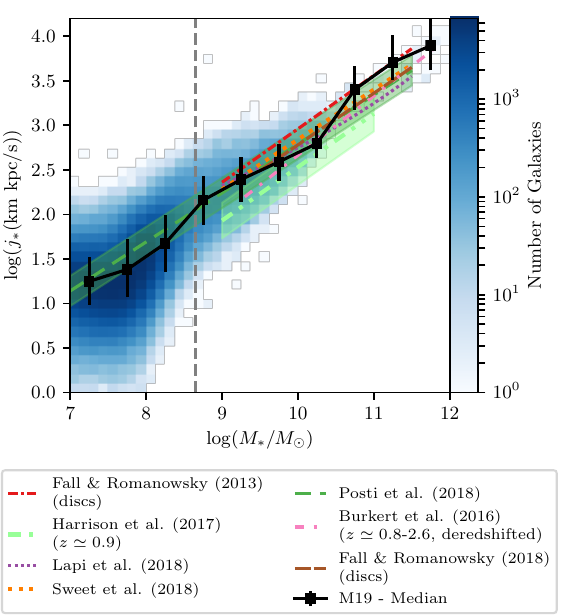}
\caption{The relation between total stellar specific angular momentum $j_\ast$ and total stellar mass $M_{\ast}$ for disc-dominated (B/T$<0.3$), central galaxies at $z=0$ in \meraxes (blue density plot). The black squares and errorbars represent the median and 16th and 84th percentile ranges of galaxy angular momenta from our model, in mass bins with width 0.5 dex. A range of observational data is also shown (see legend). The grey dotted line indicates the stellar mass below which \textit{Tiamat} and \textit{Tiamat-125-HR} are not converged (see Appendix \ref{Appendix:Resolution}).}
\label{AngularMomentumMass}
\end{center}
\end{figure}

\begin{figure*}
\begin{center}
\includegraphics[scale=0.9]{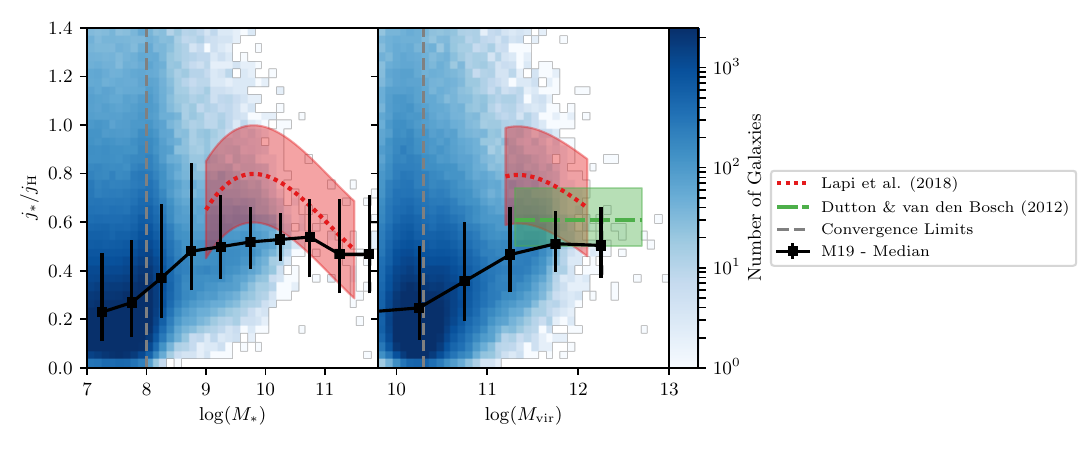}
\caption{The relation between total stellar specific angular momentum $j_\ast/j_{\textrm{DM}}$ and total stellar mass (\textit{left panel}) and host halo mass (\textit{right panel}), for disc-dominated (B/T$<0.3$), central galaxies at $z=0$ in \meraxes (blue density plot).
The black squares and errorbars represent the median and 16th and 84th percentile ranges from our model. The \citet{Lapi2018} and \citet{Dutton2012} observations are also shown (see legend). The grey dotted lines indicate the stellar mass below which \textit{Tiamat} and \textit{Tiamat-125-HR} are not converged, and the (100 particle) halo mass resolution limit for \textit{Tiamat-125-HR}.}
\label{AMRatioMass}
\end{center}
\end{figure*}

\begin{figure}
\begin{center}
\includegraphics[scale=0.9]{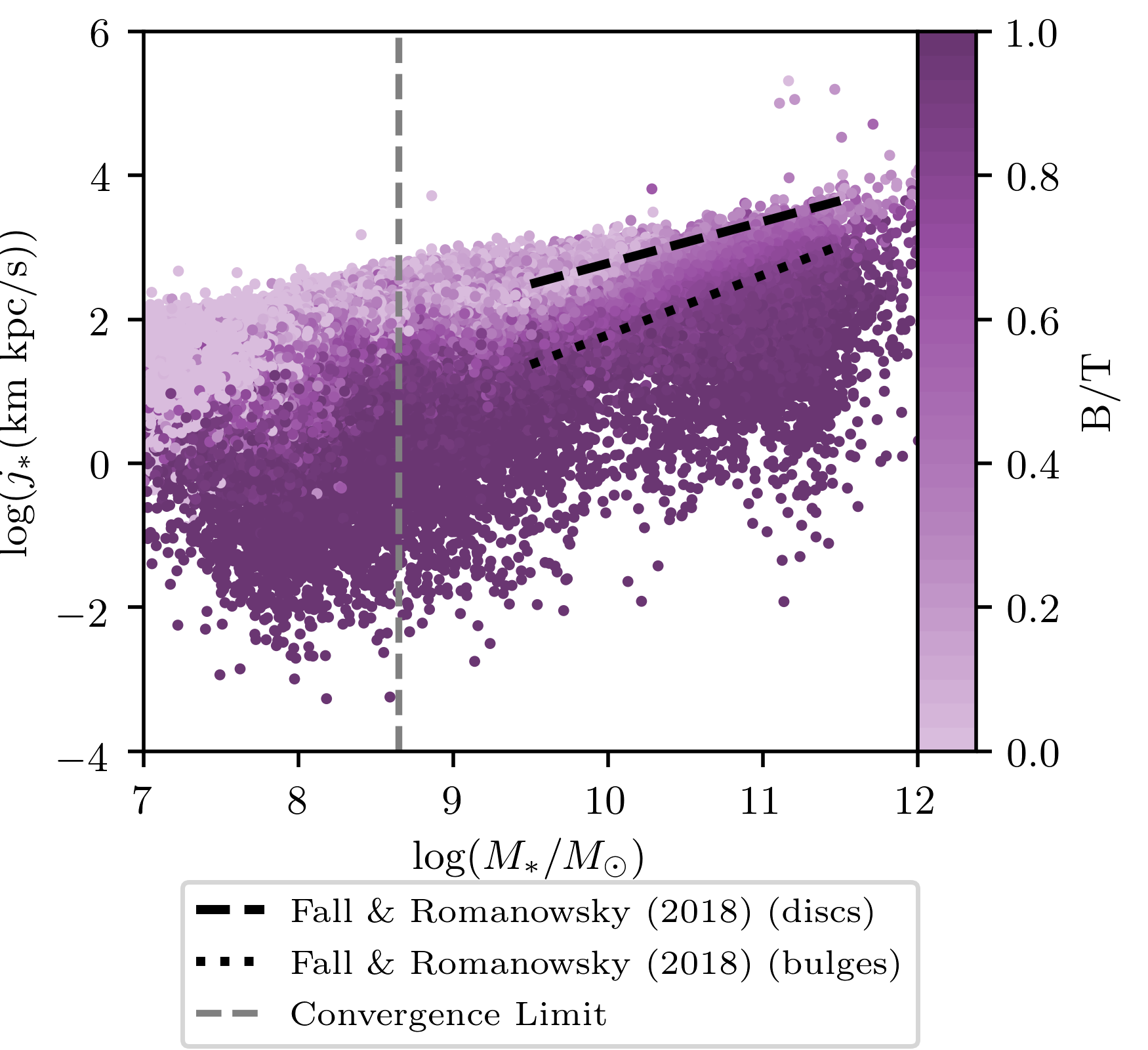}
\caption{The relation between total stellar specific angular momentum $j_\ast$ and total stellar mass $M_\ast$ for central \meraxes galaxies at $z=0$, coloured by their bulge-to-total mass ratio B/T. The grey dotted line indicates the stellar mass below which \textit{Tiamat} and \textit{Tiamat-125-HR} are not converged (see Appendix \ref{Appendix:Resolution}). Also plotted are the observed relations for discs and bulges found by \citet{Fall2018}.}
\label{AngularMomentumMass_BT}
\end{center}
\end{figure}

In this section we investigate the success of our model at reproducing the angular momentum--mass relations observed in local galaxies.
We consider only disc-dominated galaxies, with bulge-to-total mass ratio B/T$<0.3$, in order to compare with observations, which are of predominantly discy galaxies. We continue to restrict our analysis to only model galaxies classified as centrals, with $M_\ast>10^7M_\odot$. Since we assume that the bulge component has negligible angular momentum relative to the disc, we approximate the total stellar specific angular momentum as $j_\ast=J_\ast/M_\ast=J_\ast/(M_{\ast\textrm{ disc}}+M_{\ast\textrm{ bulge}})$. We refer to the stellar-disc specific angular momentum as $j_{\ast\textrm{ disc}}=J_\ast/M_{\ast\textrm{ disc}}$.

The correlation between stellar specific angular momentum $j_\ast$ and total stellar mass $M_\ast$ (commonly referred to as the Fall relation) is claimed to be one of the most fundamental galaxy scaling relations \citep[e.g.][]{Fall2018}. We show our predicted  $j_\ast$--$M_\ast$ relation at $z=0$ in Figure \ref{AngularMomentumMass}\footnote{We note that plotting our model $j_{\ast\textrm{ disc}}$--$M_{\ast\textrm{ disc}}$ relation produces a very similar relation to that in Figure \ref{AngularMomentumMass}, as we only select disc-dominated galaxies.}. We also show a range of observations of both the $j_\ast$--$M_\ast$ relation, and of the $j_{\ast\textrm{ disc}}$--$M_{\ast\textrm{ disc}}$ relation for disc galaxies \citep{Fall2013,Burkert2016,Harrison2017,Lapi2018,Sweet2018,Posti2018,Fall2018}\footnote{Note that the \citet{Fall2018} relation is an updated fit to the same data used in \citet{Fall2013}}, to which our model shows remarkable agreement.

The  \citet{Posti2018} observations extend to low stellar masses, where they find that the $j_\ast$--$M_\ast$ relation remains as a single, unbroken power-law, challenging, for example, the \citet{Stevens2016} galaxy-formation model which predicts a flattening of the relation at low masses ($M_\ast<10^9M_\odot$). 
Our model shows no significant flattening of the $j_\ast$--$M_\ast$ relation at $M_\ast<10^9M_\odot$, however, our predictions at these masses are insignificant due to our simulation convergence limits of $M_\ast=10^{8.65}M_\odot$. The lack of flattening produced by our model is consistent with the \citet{Posti2018} observations, as well as the \citet{Lagos2017} and \citet{Zoldan2018} models, for example.
\citet{Posti2018} posit that this flattening is due to assuming a constant disc-to-halo specific angular momentum ratio$j_d/j_H$ with mass, since they find that lower mass galaxies require a much smaller $j_d/j_H$.  We make no explicit assumptions regarding the $j_d/j_H$ ratio and its mass dependence; once we assume that $j_{\textrm{cold}}/j_H=1$ for gas cooling from the halo, the model then self-consistently tracks changes in the angular momentum for each galaxy throughout its history. We show the relation between $j_\ast/j_H$ ratio and stellar and halo mass for our galaxies in Figure \ref{AMRatioMass}.

The angular momentum retention factor, $j/j_H$, encapsulates how much angular momentum from the dark matter halo is retained by the galaxy as it forms and evolves. This can be affected by the many processes that alter the state of the galaxy, such as gas cooling, friction and feedback processes, each of which has a differing effect on the angular momentum of the galaxy, and which may also have dependencies on the galaxy and halo mass. The $j/j_H$ ratio gives an insight into the overall effect of these processes on a galaxy.
We predict a decreasing $j_\ast/j_H$ ratio with decreasing stellar and halo mass, as shown in Figure \ref{AMRatioMass}. While our model contains galaxies with a wide range of $j_\ast/j_H$ values, the distribution has a median lower than one, with $j_\ast/j_H\simeq 0.5$ at $M_{\textrm{vir}} \simeq 10^{12}M_\odot$ and $M_\ast\simeq 10^{10}M_\odot$. The ratio $j_\ast/j_H$ decreases towards lower halo masses, with a less significant decrease observed with stellar mass. 
Our model therefore suggests that the majority of galaxies lose angular momentum as they evolve, since $j_\ast/j_H<1$.

Our predictions overlap with the confidence intervals of the \citet{Lapi2018} and \citet{Dutton2012} observations.  \citet{Lapi2018} predict a decrease in $j_\ast/j_H$ towards lower stellar masses, consistent with our predictions, but in contrast, they predict an increase in $j_\ast/j_H$ towards lower halo masses, before a turn-over at $M_{\textrm{vir}}\simeq 10^{11.5}M_\odot$. \citet{Dutton2012} find that $j_\ast/j_H$ is roughly constant, with only a slight increase with increasing halo mass (we plot the average $j_\ast/j_H$ quoted by \citealt{Dutton2012}, as they provide no functional fit to this relation).
\citet{Fall2018} find $j_\ast/j_H\simeq 1$ for disc galaxies and $\simeq0.1$ for bulges, which are roughly constant over $10^{9.5}\lesssim M_\ast/M_\odot \lesssim10^{11.5}$; our predictions for the entire galaxy sample lie between these two values. Our results are also consistent with the \citet{Burkert2016} study which found $j_d/j_H=1.0\pm0.6$. 

We note that the model of \citet{Posti2018b} predicts that, for a range of observed stellar-to-halo mass relations, to match the $j_\ast$--$M_\ast$ relation late-type galaxies must have $j_\ast/j_H\simeq 0.3$ at $M_\ast\simeq10^{8.5}M_\odot$ or $M_{\textrm{vir}}\simeq 10^{11}M_\odot$. They find that $j_\ast/j_H$ should increase slowly with stellar mass, until a turn-over at $M_\ast\simeq10^{10.5}M_\odot$, beyond which $j_\ast/j_H$ steeply decreases. As a function of halo mass, $j_\ast/j_H$ increases with mass until $M_{\textrm{vir}}\simeq 10^{12}M_\odot$, above which it shows a clear decrease with increasing mass. The trends that \citet{Posti2018b} find at low halo and stellar masses is consistent with the findings of our simulation, however we have too few galaxies with $M_\ast>10^{10.5}M_\odot$ or $M_{\textrm{vir}}> 10^{12}M_\odot$ to investigate whether we also predict a decrease at these high masses.
We make different predictions to previous simulations and semi-analytic models, such as \citet{Pedrosa2015}, who find roughly no mass dependence of $j_d/j_H$ in their hydrodynamical simulation, as does the Dark SAGE semi-analytic model \citep{Stevens2016}, which predicts $j_\ast/j_H=0.4\pm0.29$ for spiral galaxies. The FIRE-2 simulation, however, also finds that the ratio of galaxy to halo specific angular momentum increases with stellar mass \citep{ElBadry2017}.

Finally, we consider the morphological dependence of the specific angular momentum--mass relation, as observations suggest that mass, angular momentum and bulge-to-total ratio are strongly correlated \citep[e.g.][]{Romanowsky2012,Fall2013,Obreschkow2014,Sweet2018}, with galaxies with higher B/T having lower specific angular momenta, at a fixed stellar mass \citep{Fall2013,Sweet2018,Fall2018}.
We plot the $j_\ast$--$M_\ast$ relation for galaxies in our model, coloured by their bulge-to-total mass ratio B/T, in Figure \ref{AngularMomentumMass_BT}. We see a clear trend of bulge-dominated galaxies lying below the disc-dominated sequence, consistent with the \citet{Fall2018} observations. This trend arises naturally from our assumption that the bulge component has negligible angular momentum.
The Illustris \citep{Genel2015} and Magneticum Pathfinder \citep{Teklu2015} simulations also predict a similar correlation between galaxy type and $j_\ast$, though they do not have visual morphological classifications of their galaxies. The hydrodynamical simulation of \citet{Pedrosa2015} also predicts a similar trend, with bulge and disc galaxies showing a relation with the same slopes, with the offset not dependent on the stellar or halo mass.

Overall, our model reproduces the local angular momentum--mass scaling relations well.

\section{High-redshift galaxy sizes}
\label{HighZSizes}
In Section \ref{FurtherValidation} we showed that \meraxes now predicts local galaxy sizes, angular momentum and morphologies that are in great agreement with observations. This is an important test of the galaxy formation and evolution model, and gives confidence in our predictions at higher redshifts. In the following sections, we show our predictions for the high-redshift size--luminosity and size--mass relations, alongside the redshift evolution of the median disc size \citep[for further discussion on these relations see the review by][]{Dayal2018}.

\subsection{Size--luminosity relations}
\begin{figure*}
\begin{center}
\includegraphics[scale=0.9]{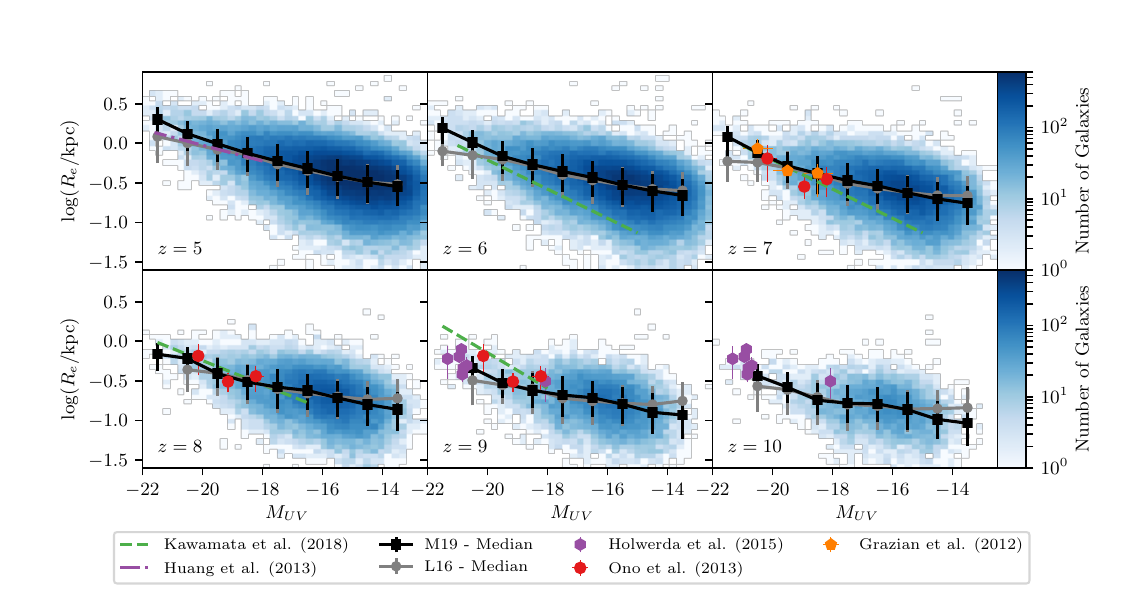}
\caption{The size--luminosity relations for our central model galaxies at $z=10$--5 (density plots), alongside the observations of \citet{Grazian2012}, \citet{Ono2013}, \citet{Huang2013}, \citet{Holwerda2015} and \citet{Kawamata2018} (see legend). The black squares (grey circles) and errorbars represent the median and 16th and 84th percentile ranges of galaxy radii from our model (the \citetalias{Liu2016} model) in bins with a width of 1 magnitude. }
\label{SizeLuminosity}
\end{center}
\end{figure*}

We investigate the relationship between the sizes of stellar discs and UV magnitude of galaxies in our model from  $z=5$--10, as shown in Figure \ref{SizeLuminosity}. Here, the UV magnitude $M_{\textrm{UV}}$ is the dust-extincted luminosity at rest-frame 1600\AA, and the radius is the effective radius of the exponential disc, $R_e$. As bulges tend to be smaller than discs, this will lead to an overprediction of the sizes of bulge-dominated galaxies in comparison to the observations. We obtain the UV luminosities of each galaxy following the method described in \citet{Liu2015}. These luminosities are for the entire galaxy; the luminosities will be decomposed into their bulge and disc components in future work. 

We find that for all UV magnitudes, brighter galaxies tend to have larger sizes. This is in agreement with the \citetalias{Liu2016} predictions, however we see a steeper slope at the brightest magnitudes.
For faint galaxies with $M_{\textrm{UV}}>-14.5$ at the highest redshifts, \citetalias{Liu2016} find that the stellar disc radius does not change significantly with luminosity; our model only shows a mild flattening in comparison. This is due to the improvements of our new disc size model. In \citetalias{Liu2016}, the galaxy radius was simply related to the halo radius. However, in \meraxes galaxies can only form in haloes above the minimum gas cooling mass; this puts an artificial limit on the minimum size a galaxy could have. By tracing the evolution of disc-sizes more physically, our galaxies sizes do not show this artificial flattening at fainter magnitudes.

For comparison, we also show the \citet{Grazian2012}, \citet{Ono2013}, \citet{Huang2013}, \citet{Holwerda2015} and \citet{Kawamata2018} observations in Figure \ref{SizeLuminosity}. Our model agrees well with the \citet{Grazian2012}, \citet{Ono2013}, \citet{Huang2013} and \citet{Holwerda2015} observations. At redshifts $z=6$, 7 and 8, the best-fit relation of \citet{Kawamata2018} agrees well at the brightest magnitudes, but our model shows a flattening in the relation at lower magnitudes that is not seen in those observations. At $z=9$, \citet{Kawamata2018} find larger sizes than our model and the  \citet{Grazian2012}, \citet{Ono2013} and \citet{Holwerda2015} observations. We note that the \citet{Grazian2012}, \citet{Huang2013}, \citet{Ono2013}, and \citet{Kawamata2018} observations include comprehensive corrections for incompleteness and biases such as cosmological surface-brightness dimming, and hence their results should not be significantly affected by selection effects. However, the \citet{Holwerda2015} observations of very luminous galaxies do not include such corrections.

The size--luminosity relation is commonly described by
\begin{equation}
R_e=R_0 \left(\frac{L_{\textrm{UV}}}{L^*_{z=3}}\right)^\beta
\end{equation}
where $R_0$ is the effective radius at $L^*_{z=3}$, and $\beta$ is the slope.
$L^\ast_{z=3}$ is the characteristic UV luminosity for $z\simeq3$ Lyman-break galaxies, which corresponds to $M_{1600}=$-21.0 \citep{Steidel1999}.
This relation can be rewritten as
\begin{equation}
\log R_e = -0.4 \beta (M_{\textrm{UV}}+21)+\log R_0.
\label{eq:SizeLuminosity}
\end{equation}
We fit this relation to our model galaxies with $M<-18$ at $z=5,6,7,8,9$ and 10, with the best-fitting values for $R_0$ and $\beta$ given in Table \ref{tab:SizeLuminosity}.
We see that the slope of the size--luminosity relation $\beta$ is roughly constant from $z=5$ to $z=10$, while the normalization $R_0$ decreases significantly.
These slopes are steeper than those predicted by \citetalias{Liu2016}, which have a median value of $\beta\simeq 0.25$ for galaxies with $M_{\textrm{UV}}<-14.5$, and increase slightly with redshift.

There is wide scatter in the $\beta$ values determined through different observations. For example, local observations of spiral galaxies have found $\beta=0.253\pm0.020$ \citep{deJong2000} and $\beta=0.321\pm0.010$ \citet{Courteau2007}. At higher redshifts, $\beta=0.22$ for $z\sim4$ and $\beta=0.25$ for $z\sim5$ \citep{Huang2013}, at $z\sim7$ $\beta=0.3$--0.5 \citep{Grazian2012} or $\beta=0.24\pm0.06$ \citep{Holwerda2015}, and \citet{Shibuya2015} find $\beta=0.27\pm0.01$ at $z\sim0$--8, with no evolution. Our values are broadly consistent with these results.

Analytically, \citet{Wyithe2011} predict that for a simple star formation model with no feedback, $\beta=1/3$. For models including supernova feedback $\beta$ decreases, to 1/4 for a model with supernova wind conserving momentum in their interaction with the galactic gas, or 1/5 if energy is instead conserved \citep{Wyithe2011}. Our predictions for $\beta$ are most consistent with the model with no supernova feedback, $\beta=1/3$. \meraxes includes a supernova feedback model which conserves energy, however, due to the complications of the hierarchical model such a comparison is not straightforward. However, we note that including fainter galaxies in the fit reduces $\beta$ to be more consistent with $\beta=1/5$.

\begin{table}
\begin{center}
\caption{The best fitting parameters for the fits to the size--mass (Equation \ref{eq:SizeMass}) and size--luminosity (Equation \ref{eq:SizeLuminosity}) relations for our model galaxies at $z=5$--10, where $R_0$ is the normalization, and $b$ and $\beta$ are the slopes of the two relations, respectively. The size--luminosity relation is fit only to galaxies with $M_{\textrm{UV}}<-18$.}
\begin{tabular}{ccccccccc}
\hline
&    \multicolumn{2}{c}{$M_\ast$} &    \multicolumn{2}{c}{$L_{\textrm{UV}}$ ($M_{\textrm{UV}}<-18$)}\\
 $z$ &     $R_0$/kpc   &     $b$    &  $R_0$/kpc  & $\beta$ \\
\hline

5  &  0.841 $\pm$ 0.006 &  0.242 $\pm$ 0.002 &   1.53 $\pm$ 0.03 &   0.32  $\pm$  0.01 \\
6  &  0.712 $\pm$ 0.008 &  0.246  $\pm$ 0.003 &   1.14  $\pm$  0.03 &   0.34  $\pm$  0.01 \\
7  &  0.616 $\pm$ 0.010 &  0.253 $\pm$ 0.004 &   0.85 $\pm$  0.02 &   0.32  $\pm$ 0.01 \\
8  &  0.545 $\pm$ 0.012 &  0.263 $\pm$ 0.006 &   0.66  $\pm$ 0.02 &   0.33  $\pm$ 0.02 \\
9  &  0.467 $\pm$ 0.015 &  0.263 $\pm$ 0.008 &   0.49  $\pm$ 0.02 &   0.32  $\pm$  0.02 \\
10 &  0.425  $\pm$ 0.016 &  0.277 $\pm$ 0.009 &   0.43  $\pm$  0.02 &   0.36  $\pm$ 0.02 \\

\hline
\label{tab:SizeLuminosity}
\end{tabular}
\end{center}
\end{table}

\subsection{Size--mass relations}

\begin{figure*}
\begin{center}
\includegraphics[scale=0.9]{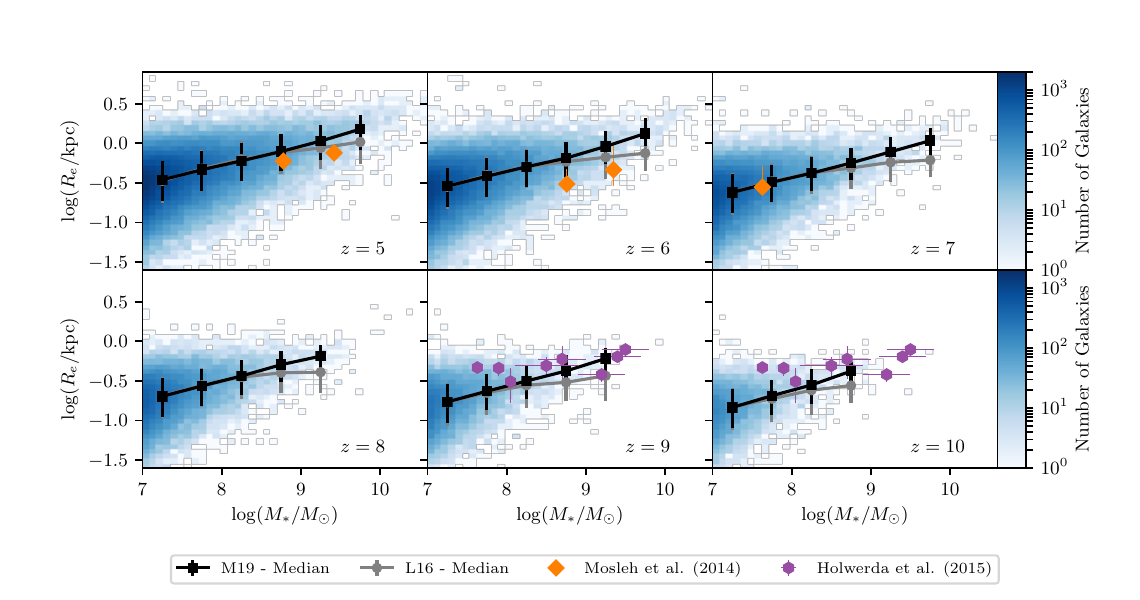}
\caption{The size--mass relations for our central model galaxies at $z=10$--5 (density plots), alongside the \citet{Mosleh2012} and \citet{Holwerda2015} observations (see legend). The black squares (grey circles) and errorbars represent the median and 16th and 84th percentile ranges of galaxy radii from our model (the \citetalias{Liu2016} model) in bins with a width of $0.5\log M_\ast$. }
\label{SizeMass}
\end{center}
\end{figure*}

The relations between the stellar disc effective radius and galaxy stellar mass from  $z=5$--10 are shown in Figure \ref{SizeMass}, for our model galaxies and the \citet{Mosleh2012} and \citet{Holwerda2015} observations. Our model shows a clear trend of increasing disc size with increasing stellar mass.
The \citet{Mosleh2012} observations match well with the model predictions at $z=5$ and 7, while being smaller than our median at $z=6$. The \citet{Holwerda2015} galaxies also match well with our predictions. We note that these observations do not correct for measurement biases, however tests by \citet{Mosleh2012} find that systematic uncertainties on their measured sizes are very small ($<10\%$). Our relations are consistent with the previous \meraxes results of \citetalias{Liu2016}, which are also shown in Figure \ref{SizeMass}.

We fit the relation
\begin{equation}
R_e=R_0 \left(\frac{M_\ast}{M_0}\right)^b
\label{eq:SizeMass}
\end{equation}
where  $M_0 =10^9 M_\odot$, as in \citet{Holwerda2015}, to our model galaxies at $z=5,6,7,8,9$ and 10, with the best-fitting values for $R_0$ and $b$ given in Table \ref{tab:SizeLuminosity}.  The normalization $R_0$ decreases from $z=5$ to $z=10$, while the slope $b$ increases. Our values for $b$ are consistent with some of those derived by \citet{Holwerda2015}, though their values vary significantly depending on the observations used; they derive $b=0.14\pm0.20$ for $z=6$ (from the \citealt{Mosleh2012} data), $b=0.24\pm0.08$ and $b=1.35\pm0.34$ for $z=7$ (from the \citealt{Grazian2012} and \citealt{Ono2013} data, respectively), and $b=0.12\pm0.06$ at $z=9$--10 (from the \citealt{Holwerda2015} data).

\subsection{Redshift evolution of disc size}
\begin{figure}
\begin{center}
\includegraphics[scale=0.9]{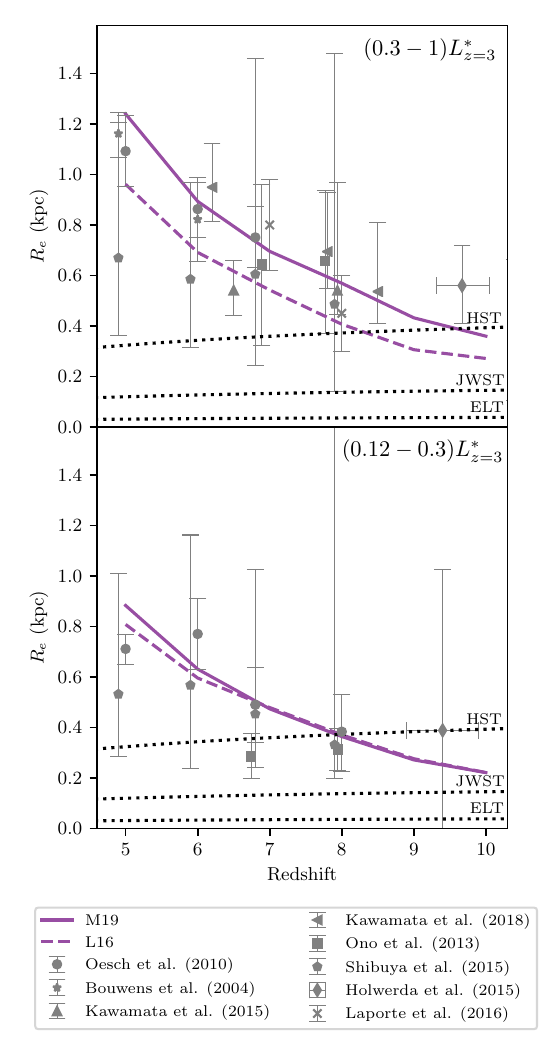}
\caption{The redshift evolution of the median galaxy effective radius for central galaxies in the luminosity range (0.3--1)$L^\ast_{z=3}$ (\textit{upper panel}) and (0.12--0.3)$L^\ast_{z=3}$ (\textit{lower panel}). Our \meraxes predictions are shown for all central galaxies (solid purple lines), and the predictions by the \citetalias{Liu2016} \meraxes model are also shown (dashed purple lines). For these simulated galaxies we only plot bins that contain 10 or more galaxies. A range of observations and their errors are also shown (see legend), as are the resolution limits for HST, JWST and GMT (black dotted lines).}
\label{SizeRedshift}
\end{center}
\end{figure}

We show the predictions of our model for the redshift evolution of the stellar disc effective radius in Figure \ref{SizeRedshift}. We split our galaxies into two luminosity bins, (0.12--0.3) $L^\ast_{z=3}$ and (0.3--1) $L^\ast_{z=3}$, in order to compare with observations; these luminosity ranges correspond to UV magnitudes -18.7 to -19.7 and -19.7 to -21.0, respectively. The observations of \citet{Bouwens2004}, \citet{Oesch2010}, \citet{Ono2013}, \citet{Kawamata2015}, \citet{Shibuya2015}, \citet{Holwerda2015}, \citet{Laporte2016} and \citet{Kawamata2018} are also shown. Our predictions are consistent with these observations, but are somewhat steeper.

We note that the \citet{Oesch2010}, \citet{Ono2013}, \citet{Kawamata2015}, and \citet{Kawamata2018} observations include comprehensive corrections for incompleteness and biases such as cosmological surface-brightness dimming, and hence their results should not be significantly affected by selection effects. \citet{Bouwens2004} and \citet{Shibuya2015}, however, find that surface brightness biases have no significant effect on their measured galaxy sizes, and are important only for galaxies close to the magnitude limit. Hence, observations that do not account for this should not significantly underestimate the true galaxy size. 

The steepness of our model relative to the observations can be seen in the fitting of the relation with the form $R_e \propto (1+z)^{-m}$, the functional form commonly used to describe the size-redshift relation. We obtain this $m$ by performing non-linear least squares on the sizes of each model galaxy at $z=5,6,7,8,9$ and 10.
We obtain $m=1.98\pm0.07$ for galaxies with (0.3--1)$L^\ast_{z=3}$, with $R_e(z=7)=0.74\pm0.01$, and $m=2.15\pm0.05$ for galaxies with (0.12--0.3)$L^\ast_{z=3}$, with $R_e(z=7)=0.522\pm0.006$. 
These are within 3$\sigma$ of the \citetalias{Liu2016} values, of $m=2.00\pm0.07$ for galaxies with (0.3--1)$L^\ast_{z=3}$ and $m=2.02\pm0.04$ for galaxies with (0.12--0.3)$L^\ast_{z=3}$
Our predictions are also consistent with the FIRE-2 hydrodynamical simulation \citep{Ma2018}, which predicts $m=1$--2, depending on the fixed mass or luminosity at which the relation is calculated.
Our predictions for $m$ are higher than those derived from the individual observational data sets shown in Figure \ref{SizeRedshift}, which all predict $1\lesssim m \lesssim 1.5$ (except for \citet{Holwerda2015}, who predict $m=0.76\pm0.12$). 
However, \citetalias{Liu2016} show that a combined fit to these $z>5$ observations produces larger values of $m$ than the individual analyses, which include $z<5$ data, of $m=1.64\pm0.30$ and $m=1.82\pm0.51$ for the (0.3--1)$L^\ast_{z=3}$ and (0.12--0.3)$L^\ast_{z=3}$ ranges, respectively; these are consistent with our predictions at $<2\sigma$. 
Analytically, models that assume that galaxy size is proportional to $R_{\textrm{vir}}$ \citep[e.g.][]{Mo1998} predict evolution with an index of $m=1$ for fixed halo mass, or $m=1.5$ for fixed circular velocity. However, these are simplistic models that assume no star-formation feedback. Including physical prescriptions for feedback processes alters a model's predictions for $m$ \citep{Wyithe2011}.

In Figure \ref{SizeRedshift}, we show the resolution limits of HST, the James Webb Space Telescope (JWST), and the Giant Magellan Telescope (GMT). For resolution of a telescope with mirror diameter $D_{\textrm{tel}}$ is
\begin{equation}
\Delta l = \Delta \Omega d_A = \frac{1.22\lambda}{D_{\textrm{tel}}} d_A
\end{equation}
where  $\Delta \Omega$ is the angular resolution determined by the Rayleigh criterion, $d_A$ is the angular diameter distance,  and $\lambda=1600(1+z)$ \AA~ is the observed wavelength of UV photons. 
Studies using HST indicate that galaxies can be resolved if $R_e \gtrsim \Delta l / 2$ \citep[e.g.][]{Ono2013,Shibuya2015} and so we adopt the resolution limit $R_{\textrm{min}}=\Delta l / 2$. 

For HST ($D_{\textrm{tel}}=2.4$m), we find that the typical (0.12--0.3)$L^\ast_{z=3}$ galaxy can be resolved to $z\sim8$, and the typical (0.3--1)$L^\ast_{z=3}$ galaxy to $z\sim9$.
With its larger mirrors, JWST ($D_{\textrm{tel}}=6.5$m) can resolve the majority of these (0.12--1)$L^\ast_{z=3}$ galaxies at $z=10$.
The next generation of ground-based telescopes such as the GMT ($D_{\textrm{tel}}=25$m) will be able to resolve at least 98 per cent of all $z=10$ galaxies.

\section{Redshift evolution of angular momentum}
\label{HighZAngularMomentum}

\begin{figure*}
\begin{center}
\includegraphics[scale=0.9]{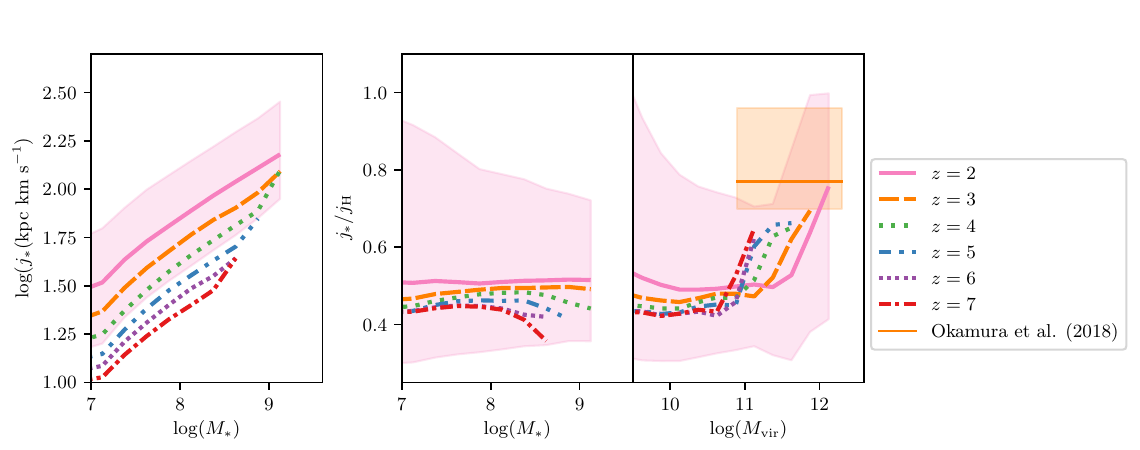}
\caption{\textit{Left panel:} The median relation between total stellar specific angular momentum $j_\ast$ and total stellar mass $M_{\ast}$ for disc-dominated (B/T$<0.3$), central galaxies at $z=2$--7 in \meraxes (see legend).
\textit{Middle and Right panels:} The median ratio of stellar to halo specific angular momentum $j_\ast/j_{\textrm{DM}}$ as a function of total stellar mass and halo mass, for disc-dominated (B/T$<0.3$), central galaxies at $z=2$--7 in \meraxes (see legend).
The pink shaded region shows the 16th and 84th percentile range for the $z=2$ model galaxies.
The \citet{Okamura2018} observations are also shown (orange solid line, 1$\sigma$ errors are shaded).}
\label{AngularMomentumEvolution}
\end{center}
\end{figure*}

In Section \ref{FurtherValidation} we showed that \meraxes can reproduce the observed local relations between angular momentum and mass. We now present the model's predictions for the evolution of these relations with redshift.
We restrict our analysis to disc-dominated (B/T$<0.3$), central galaxies.

We plot the median total stellar specific angular momentum as a function of total stellar mass from $z=7$ to $z=2$ in Figure \ref{AngularMomentumEvolution}. The relation evolves to higher values of $j_\ast$ for lower redshifts, at a fixed stellar mass. From $z=7$ to $z=2$, the relation increases by $\sim0.5$ dex, significant relative to the 16th--84th percentile range for the $z=2$ relation.

At lower redshifts, observations show a range of results for the evolution of the $j_\ast$--$M_\ast$ relation.
For example, \citet{Contini2016} investigated a sample of low-mass star-forming galaxies at $0.2<z<1.4$ (median $z\simeq0.6$) and found them to follow the $z=0$ relation. 
The $z=1$ observations of \citet{Marasco2019} are also consistent with no evolution.
\citet{Harrison2017}, however, find that star forming galaxies at $z\simeq0.9$ have 0.2--0.3 dex less angular momentum for a fixed stellar mass compared to $z=0$ galaxies.
\citet{Swinbank2017} study star-forming galaxies in $z=0.23$--1.65 and find that most massive star-forming discs at $z\simeq0$ must have increased their specific angular momentum by around a factor of 3 since $z\simeq1$.
 \citet{Alcorn2018} find a decrease in angular momentum for star-forming galaxies at $2<z<2.5$ consistent with 
\citet{Harrison2017}, however they find little to no increase in the $j_\ast$--$M_\ast$ relation if its slope is constrained to 2/3.

Our model is consistent with the predictions of Dark SAGE \citep{Stevens2016}, which investigates redshifts $0<z<4.8$ and finds a weak trend for galaxies of fixed mass to have lower specific angular momentum at higher redshifts.
In contrast, the hydrodynamical simulation of \citet{Pedrosa2015} finds that the $j_d$--$M_\ast$ relations are statistically unchanged up to $z\simeq2$, and their ratio $j_d/j_H\sim1$ shows no clear evolution with redshift. \citet{Zoldan2019} also predict with their semi-analytic model that the $j_\ast$--$M_\ast$ relation shows little evolution with redshift to $z\sim2$ for star-forming galaxies, however quiescent and early-type galaxies show a moderately decreasing $j_\ast$ with increasing redshift.

We also plot the relations between $j_\ast/j_H$ and total stellar mass and halo mass at high redshifts in Figure \ref{AngularMomentumEvolution}.
The $j_\ast/j_H$--$M_{\ast}$ relation shows an increase in the $j_\ast/j_H$ ratio with decreasing redshift at constant stellar mass, increasing by around $\sim0.1$ for $10^{7.5}<M_{\ast}/M_\odot<10^{9}$  from $z=7$ to $z=2$, much less than the scatter in the relation.
The $j_\ast/j_H$--$M_{\textrm{vir}}$ relation also shows no significant redshift evolution.

\citet{Okamura2018} study the angular momentum of galaxies at $z=2,3,4$ by using the \citet{Mo1998} analytic model to estimate $j_{\ast\textrm{ disc}}$ from the measured galaxy sizes. They use two methods to estimate the halo masses, finding that $j_{\ast\textrm{ disc}}/j_H=0.77\pm0.06$ from clustering analysis and $j_{\ast\textrm{ disc}}/j_H=0.83\pm0.13$ from abundance matching, with no strong redshift evolution. They find a weak dependence of $j_{\ast\textrm{ disc}}/j_H$ on $M_{\textrm{vir}}$, with an increase in $j_{\ast\textrm{ disc}}/j_H$ with decreasing stellar mass. This is in contrast to our predictions. We note, however, that their halo masses are strongly dependent on the estimation method used, and since they adopt an analytic model to estimate $j_{\ast\textrm{ disc}}$ instead of measuring it kinematically, there may be uncertainty in this mass trend.

\section{High-redshift galaxy morphologies}
\label{HighZMorphology}

\begin{figure*}
\begin{center}
\includegraphics[scale=0.9]{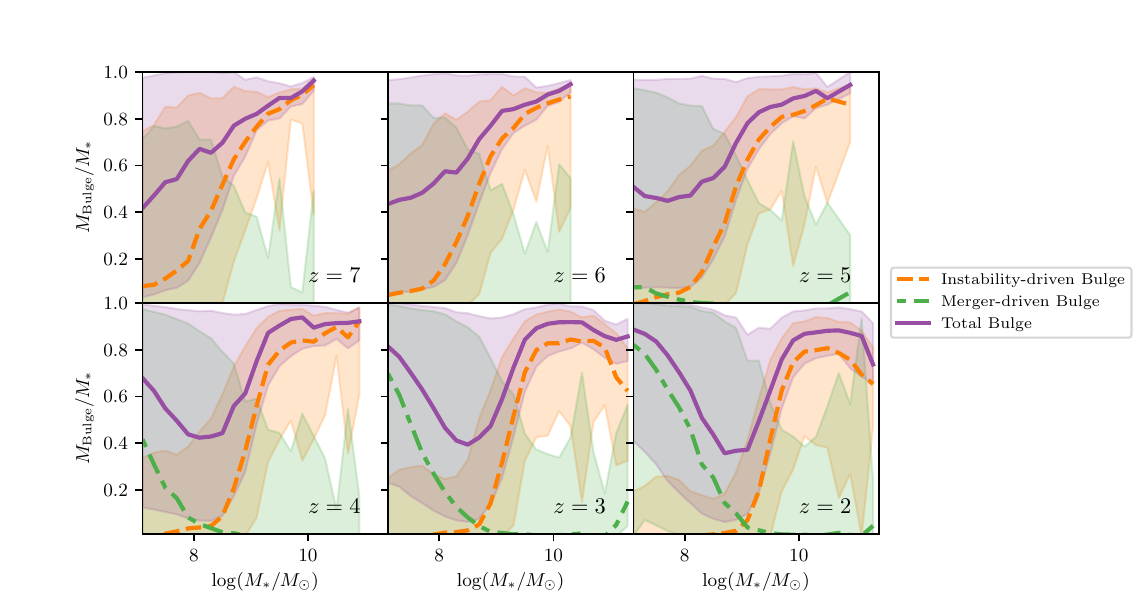}
\caption{
The fraction of stellar mass contained in the (combined) bulge components of model galaxies as a function of total stellar mass at $z=2$--7. Only galaxies classified as centrals are shown.
Purple solid lines show the median galaxy bulge fraction in mass bins of width 0.2 dex, and the purple shaded regions show the 16th and 84th percentile range.
Also shown are the median mass fractions of the instability-driven bulge (orange dashed) and merger-driven bulge (green dot-dashed), alongside their 16th and 84th percentile ranges.}
\label{fig:BulgeFractionEvolution}
\end{center}
\end{figure*}

In Section \ref{FurtherValidation} we showed that \meraxes can reasonably reproduce the morphological distribution of observed low-redshift galaxies. Here we show the \meraxes predictions for galaxy morphologies at high-redshift.

In Figure \ref{fig:BulgeFractionEvolution} we show the redshift evolution of the B/T ratio as a function of total stellar mass. 
As redshift increases, the dip in the B/T ratio becomes shallower and shifts to lower masses. This is a result of an evolution of the masses at which the merger- and instability-driven growth modes are effective. The masses at which the merger-driven mode typically operates shift to lower values at higher redshift, and the steep cut-off between the masses where the instability-driven mode does and does not operate flattens at higher redshifts. This results in an average B/T that is  higher at high redshifts, suggesting that high-redshift galaxies are more likely to be bulge-dominated than those at low redshift.

This is consistent with the current observations, which suggest that the spheroidal fraction increases to higher redshifts \citep[e.g.][]{Ravindranath2006,Lotz2006,Dahlen2007,Shibuya2015}. This is supported by the IllustrisTNG hydrosimulation \citep{Pillepich2019}. The EAGLE hydrosimulation \citep{Trayford2018} finds that the both the disc and bulge fractions decrease to higher redshifts, with the high-redshift population dominated instead by asymmetric galaxies.

\section{Conclusions}
\label{Conclusion}
In this paper we introduce updates to the semi-analytic model \meraxes in order to investigate the evolution of sizes, angular momenta and morphologies of galaxies to high redshifts. These updates include a prescription for bulge formation and growth with bulges formed through both galaxy mergers and disc instabilities, tracking of the angular momentum of the gas and stellar discs throughout their evolution, and a new way of determining galaxy disc sizes and velocities incrementally as they evolve. The model is calibrated to the observed stellar mass functions at $z=8$--0 and the black hole--bulge mass relation at $z=0$. 

At low redshifts, the model reproduces the observed galaxy size--mass relation well. We produce galaxy morphologies that are consistent with observations, with galaxies at $M_\ast\simeq10^{9.5}M_\odot$ forming their bulges predominantly through galaxy mergers, while disc instabilities dominate for bulge growth in galaxies with $M_\ast\simeq10^{10}$--$10^{11.5}M_\odot$, and major mergers form massive ellipticals for galaxies with $M_\ast>10^{11.5}M_\odot$.
We predict a specific angular momentum--mass relation that is consistent with observations, and shows no flattening at low-masses, thus being more consistent with observations than previous semi-analytic models \citep[e.g.][]{Stevens2016}. We find that the specific angular momentum--mass relation depends on galaxy morphology, with bulge dominated galaxies lying below the disc-dominated sequence, consistent with the observed trends.
We predict that the ratio between stellar and halo specific angular momentum is typically less than one, suggesting that galaxies lose angular momentum as they evolve, and that it decreases with halo and stellar mass.

At high redshifts, we make the following predictions:
\begin{itemize}
\item The size--luminosity relation $R_e=R_0 (L_{\textrm{UV}}/L^*_{z=3})^\beta$ has $\beta=0.32\pm0.01$ and $R_0=1.53\pm0.03$ kpc at $z=5$, with the normalisation decreasing at higher redshifts but the slope remaining roughly constant, and is consistent with available high-redshift observations.
\item The size--mass relation $R_e=R_0 \left(\frac{M_\ast}{M_0}\right)^b$ has $b=0.242\pm0.002$ and $R_0=0.841\pm0.006$ kpc at $z=5$, with both slope and normalisation decreasing at higher redshifts, and is reasonably consistent with the available observations.
\item The median size of a galaxy disc decreases with redshift as $R_e \propto (1+z)^{-m}$, where  $m=1.98\pm0.07$ for galaxies with (0.3--1)$L^\ast_{z=3}$, with $R_e(z=7)=0.74\pm0.01$, and $m=2.15\pm0.05$ for galaxies with (0.12--0.3)$L^\ast_{z=3}$, with $R_e(z=7)=0.522\pm0.006$. 
\item The specific angular momentum--stellar mass relation evolves to higher values of $j_\ast$ for lower redshifts, at a fixed stellar mass, with an increase of  $\sim0.5$ dex from $z=7$ to $z=2$.
\item The relation between the ratio of stellar and halo specific angular momentum $j_\ast/j_H$ to stellar mass shows a slight but insignificant increase with decreasing redshift.
\item Galaxies at high redshifts are predominantly bulge-dominated.
\end{itemize}

\meraxes can now predict a wide-range of relations that can be observed and tested by the next generation of telescopes, such as JWST.
For example, we are able to make predictions for the high-redshift black hole--bulge relation, which is the topic of our next paper.

\section*{Acknowledgements}
We thank the anonymous referee for their detailed comments and suggestions. We are also grateful to Michael Fall and Lorenzo Posti for their useful comments.
This research was supported by the Australian Research Council Centre of Excellence for All Sky Astrophysics in 3 Dimensions (ASTRO 3D), through project number CE170100013.
This work was performed on the OzSTAR national facility at Swinburne University of Technology. OzSTAR is funded by Swinburne University of Technology and the National Collaborative Research Infrastructure Strategy (NCRIS).
MAM acknowledges the support of an Australian Government Research Training Program (RTP) Scholarship.



\bibliography{Paper_Sizes_Arxiv2.bib} 

\appendix

\section{Resolution limits of Tiamat and Tiamat-125-HR}
\label{Appendix:Resolution}

\begin{figure*}
\begin{center}
\includegraphics[scale=0.9]{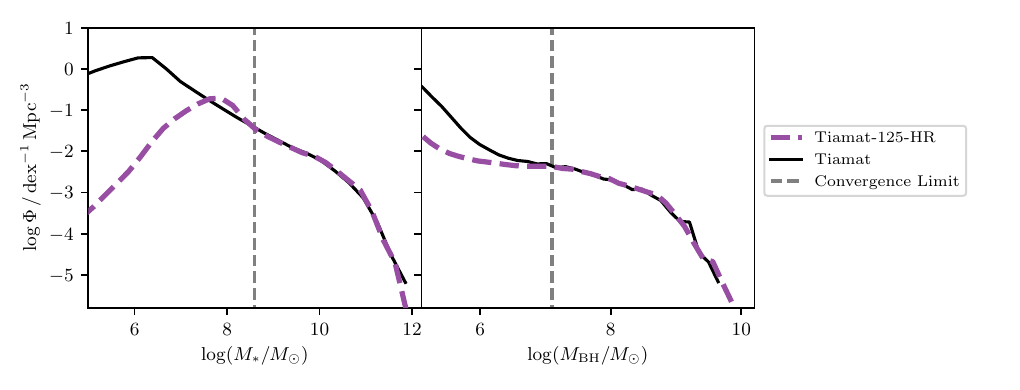}
\caption{The $z=2$ stellar mass and black hole mass functions for \meraxes applied to both \textit{Tiamat} (black solid) and \textit{Tiamat-125-HR} (purple dashed) }
\label{fig:Resolution}
\end{center}
\end{figure*}

In Figure \ref{fig:Resolution} we show the $z=2$ stellar mass, and black hole mass functions for \meraxes applied to both \textit{Tiamat} and \textit{Tiamat-125-HR}. For the stellar mass functions, we see a divergence of the \textit{Tiamat-125-HR} function from that of \textit{Tiamat} at $M_\ast=10^{8.6}M_\odot$, below which it experiences a slight upturn before a rapid turn-over. This is a signature of the resolution of the simulation. A similar divergence is also present in the black hole mass function. We use these signatures to determine the simulation resolution as follows.
We choose the mass at which the stellar mass function begins to upturn as the stellar mass resolution; at $z=0$, this corresponds to a mass of 
$M_\ast=10^{8.65} M_\odot$. For the black hole mass function, we take mass resolution as the mass at which the black hole mass function begins to flatten---$M_{\textrm{BH}}=10^{7.35}M_\odot$ at $z=0$. 

\section{UV luminosities and star formation rates}
\label{Appendix:Validation}

\begin{figure*}
\begin{center}
\includegraphics[scale=0.9]{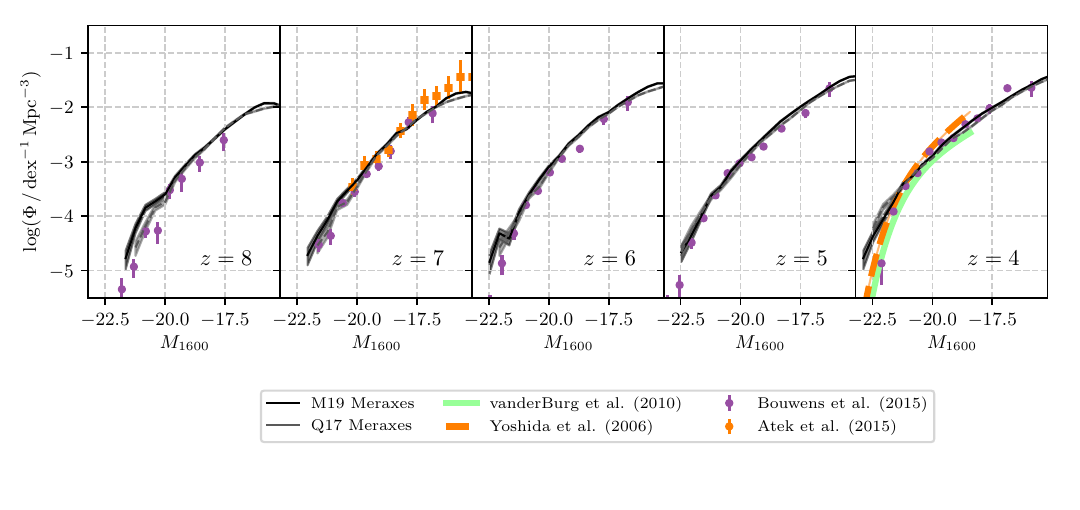}
\caption{Galaxy luminosity functions at $z=8$--4 from the \citetalias{Qin2017} \meraxes and our new model compared to the observations of \citet{Yoshida2006,VanDerBurg2010,Bouwens2015,Atek2015} (see legend).}
\label{GLF}
\end{center}
\end{figure*}

The UV luminosity functions at $z=8$--4 are shown in Figure \ref{GLF}, and agree with both those from the \citetalias{Qin2017} \meraxes and observations, but diverge slightly at the faint end.

\begin{figure*}
\begin{center}
\includegraphics[scale=0.9]{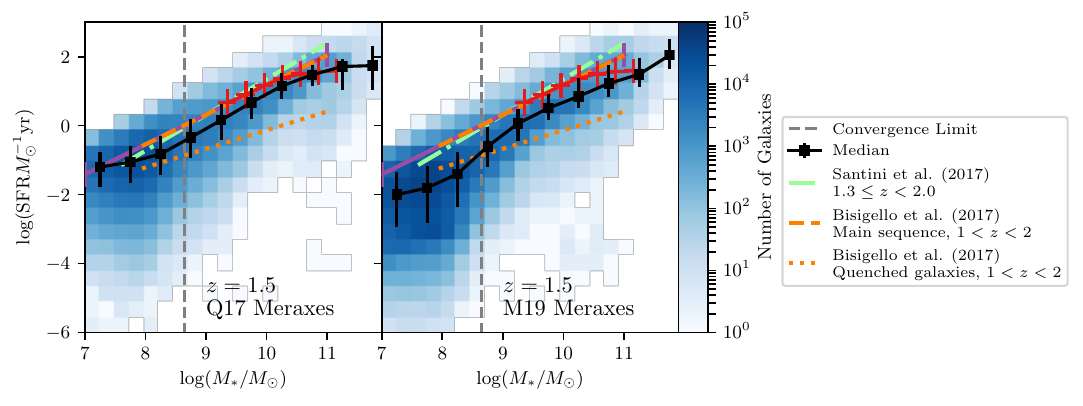}
\caption{The relation between star formation rate (SFR) and stellar mass ($M_\ast$) at $z=1.5$ for both the \citetalias{Qin2017} and our updated \meraxes (M19), alongside observations \citep[][see legend]{Lee2015,Kurczynski2016,Santini2017,Bisigello2017}. The black squares and errorbars represent the median and 16th and 84th percentile ranges of the models. Only galaxies classified as centrals are shown. The grey dotted line indicates the stellar mass below which \textit{Tiamat} and \textit{Tiamat-125-HR} are not converged (see Appendix \ref{Appendix:Resolution}). }
\label{MassSFR}
\end{center}
\end{figure*}

The $\textrm{SFR}$--$M_\ast$ relations for the \citetalias{Qin2017} and new models are shown in Figure \ref{MassSFR}, alongside observations by \citet{Lee2015}, \citet{Kurczynski2016}, \citet{Santini2017} and \citet{Bisigello2017}. Both the original and new \meraxes models predict a star formation main sequence consistent with the observations, with the new \meraxes showing a relation that extends to lower star formation rates.

\bsp	
\label{lastpage}
\end{document}